\documentclass[useAMS,usenatbib]{mn2e}

\usepackage{times,epsfig}
\usepackage{graphics,graphicx}
\usepackage{amsmath}
\usepackage{amsfonts}
\usepackage{epsfig}
\usepackage{url}
\usepackage{tabularx}
\usepackage{afterpage}

\usepackage{mathrsfs}
\usepackage{amssymb}
\usepackage{bm}

\newcommand{\prd}{Physical Review D}
\newcommand{\ud}{\mathrm{d}} 
\def\lsim{\,\lower2truept\hbox{${<\atop\hbox{\raise4truept\hbox{$\sim$}}}$}\,}
\def\gsim{\,\lower2truept\hbox{${> \atop\hbox{\raise4truept\hbox{$\sim$}}}$}\,}
\def\simlt{\mathrel{\rlap{\lower 3pt\hbox{$\sim$}}
        \raise 2.0pt\hbox{$<$}}}
\def\simgt{\mathrel{\rlap{\lower 3pt\hbox{$\sim$}}
        \raise 2.0pt\hbox{$>$}}}

\def\be{\begin{equation}}
\def\ee{\end{equation}}
\def\ba{\begin{eqnarray}}
\def\ea{\end{eqnarray}}

\def\DHLhksqrt#1#2{\setbox0=\hbox{$#1\sqrt{#2\,}$}\dimen0=\ht0
\advance\dimen0-0.2\ht0
\setbox2=\hbox{\vrule height\ht0 depth -\dimen0}%
{\box0\lower0.4pt\box2}}

\newcommand{\lc}{$\Lambda$CDM }
\newcommand{\lcgr}{$\Lambda$CDM+GR }
\newcommand{\gam}{$\gamma$ }
\newcommand{\cinft}{$c_{\infty}$ }
\newcommand{\cinf}{c_{\infty} }

\title[Testing Gravity using RSD]{Testing Gravity Using Large-Scale Redshift-Space Distortions}
\author[Raccanelli et al.]{\parbox[t]{\textwidth}{Alvise Raccanelli$^{1,2}$, Daniele Bertacca$^{3,4}$, Davide Pietrobon$^{1}$, Fabian Schmidt$^{2}$, Lado Samushia$^{6}$, Nicola Bartolo$^{4,5}$, Olivier Dor\'{e}$^{1,2}$, Sabino Matarrese$^{4,5}$, Will J. Percival$^{6}$}
\vspace*{8pt}\ \\
$^{1}$ Jet Propulsion Laboratory, California Institute of Technology, Pasadena, CA 91109, USA \\
$^{2}$ California Institute of Technology, Pasadena, CA 91125, USA \\
$^{3}$ Physics Department, University of the Western Cape, Cape Town 7535, South Africa \\
$^{4}$ Dipartimento di Fisica e Astronomia ``G. GalileiÓ, Universit\`{a}  degli Studi di Padova,  via F. Marzolo, 8 I-35131 Padova, Italy\\
$^{5}$ INFN Sezione di Padova, via F. Marzolo, 8 I-35131 Padova, Italy\\
$^{6}$ Institute of Cosmology \& Gravitation, University of Portsmouth, Dennis Sciama building, Portsmouth, P01 3FX, UK
}
\date{}
\begin{document}
\maketitle
\begin{abstract}
We use Luminous Red Galaxies from the Sloan Digital Sky Survey II to test the cosmological structure growth in two alternatives to the standard $\Lambda$CDM+GR cosmological model.
We compare observed three-dimensional clustering in SDSS DR7 with theoretical predictions for the standard vanilla $\Lambda$CDM+GR model, Unified Dark Matter cosmologies and the normal branch DGP.
In computing the expected correlations in UDM cosmologies, we derive a parameterized formula for the growth factor in these models.
For our analysis we apply the methodology tested in~\cite{raccanelli10} and use the measurements of~\cite{samushia11}, that account for survey geometry, non-linear and wide-angle effects and the distribution of pair orientation.
We show that the estimate of the growth rate is potentially degenerate with wide-angle effects, meaning that extremely accurate measurements of the growth rate on large scales will need to take such effects into account.
We use measurements of the zeroth and second order moments of the correlation function from SDSS DR7 data and the Large Suite of Dark Matter Simulations (LasDamas: \citealt{mcbride11}), and perform a likelihood analysis to constrain the parameters of the models.
Using information on the clustering up to $r_{\rm max}$ = 120 Mpc/h, and after marginalizing over the bias, we find, for UDM models, a speed of sound \cinft $\le$ 6.1e-4, and, for the nDGP model, a cross-over scale $r_c \ge$ 340 Mpc, at 95\% confidence level.
\end{abstract}

\begin{keywords}
large scale structure of the Universe ---
cosmological parameters --- 
cosmology: observations --- methods : analytical
\end{keywords}


\section{Introduction}
The strangest feature of our current cosmological model is the observation that the expansion rate of the Universe is accelerating~\citep{perlmutter99,riess98}. Understanding the cause of cosmic acceleration is one of the great challenges of physics. It has been speculated that the cause of this acceleration is a cosmological constant, or perhaps some novel form of matter; our ignorance is summarized by the simple name for the cause of the observed phenomenon: Òdark energyÓ. Alternatively, it could be explained by the breakdown of Einstein's General Relativity (GR) theory of gravitation on cosmological scales (see~\citealt{durrer08} for a review on different dark energy and modified gravity models). Observations of the large-scale structure of the Universe have played an important role in developing our standard cosmological model and will play an essential role in our investigations of the origin of cosmic acceleration.

We will illustrate how it is possible to test GR and alternative models of gravity using Luminous Red Galaxies (LRG) from the Sloan Digital Sky Survey (SDSS) DR7 data. To estimate the statistical errors on our measurements we use galaxy catalogues from the Large Suite of Dark Matter Simulations (LasDamas: \citealt{mcbride11})
\footnote{http://lss.phy.vanderbilt.edu/lasdamas/},
that are designed to model the clustering of Sloan Digital Sky Survey (SDSS) galaxies.

The presence of a dark energy component in the energy-density of the Universe (or the fact that our theory of gravity needs to be modified on large scales), modifies the gravitational growth of large-scale structures. The large-scale structure we see traced by the distribution of galaxies arises through gravitational instability, which amplifies primordial fluctuations that originated in the very early Universe; the rate at which structure grows from small perturbations offers a key discriminant between cosmological models, as different models predict measurable differences in the growth rate of large-scale structure with cosmic time (e.g. \citealt{jain07, song09koyama, song09percival}). For instance, dark energy models in which general relativity is unmodified predict different large-scale structure formation compared to Modified Gravity models with the same background expansion (e.g. \citealt{DGP, carroll04, brans05, nesseris08, yamamoto08, yamamoto10}).

Observations of Redshift Space Distortions (RSD) in spectroscopic galaxy surveys are a promising way to study the pattern and the evolution of the Large Scale Structure of the Universe (\citealt{kaiser87}, \citealt{hamilton98}), as they provide constraints on the amplitude of peculiar velocities induced by structure growth, thereby allowing tests of the theory of gravity governing the growth of those perturbations.
RSD have been measured using techniques based on both correlation functions and power-spectra \citep{peacock01, hawkins03, percival04, pope04, zehavi05, tegmark06, okumura08, cabre09, guzzo08, blake11rsd}; the most recent analyses come from the BOSS DR9 catalogue~\citep{reid12, Sanchez12}.

A key element of RSD is that the motion of galaxies is independent of their properties and of the bias, that relates the baryonic matter to the total mass; therefore, measurements of peculiar velocity directly probe the matter distribution.
They also are complementary to other probes, since they depend on temporal metric perturbations, while e.g. weak lensing depends on the sum of the temporal and spatial metric perturbations and the Integrated Sachs-Wolfe effect depends on the sum of their derivatives.

The standard analysis of RSD makes use of the so-called Kaiser formalism, that relies on some assumptions, including considering only the linear regime and the distant observer approximation, and restrains the range of usable scales to 30-60 Mpc/h. There have been several attempts to model smaller scales RSD, exploring the quasi-linear regime (e.g. \citealp{scoccimarro04,taruya10,reid11,kwan11}); recently~\cite{bertaccaGR} developed a formalism to compute the correlation function including GR corrections, that arise when probing scales comparable to the Hubble scales.

In this paper we show how precise measurements of the clustering of galaxies can be used to test cosmological models;
we make use of the wide-angle methodology as tested in~\cite{raccanelli10}, that drops the distant observer approximation, combined with prescriptions and measurements of SDSS-II data from~\cite{samushia11}, to constrain two interesting alternatives to the standard cosmology: a particular class of Unified Dark Matter (UDM) models~\citep{Bertacca:2008uf, Bertacca:2010ct} and the normal-branch Dvali-Gabadadze-Porrati (DGP)~\citep{Schmidt09}.
In the process, we also derive a parameterized formula for the growth factor in UDM models, and we show that wide-angle corrections are degenerate with variations of the rate of the growth of structures, demonstrating the need to include them if one wants to measure the growth rate at percent level.

The paper is organised as follows:
in Section~\ref{sec:rsd} we briefly review the theory of RSD; 
in Section~\ref{sec:sdss} we present the catalog used; the methodology used to perform measurements is reviewed in Appendix~\ref{sec:appmethod};
in Section~\ref{sec:growth} we introduce the parameterisation of structure growth we apply in our tests and discuss degeneracies that arise from a more comprehensive description of the data; 
after discussing the theoretical growth of structures for the non-standard cosmological models,  
we present the measurements in Section~\ref{sec:res}, and set limits on the parameters in Section~\ref{sec:results}; finally, in 
Section~\ref{sec:disc}, we conclude and discuss our results.


\section{Redshift-Space Distortions}
\label{sec:rsd}
RSD arise because we infer galaxy distances from their redshifts using the Hubble law: the radial component of the peculiar velocity of individual galaxies will contribute to each redshift and will be misinterpreted as being cosmological in origin, thus altering our estimate of the distances to them. 
The relation between the redshift-space position $\mathbf{s}$ and real-space position $\mathbf{r}$  is:
\begin{equation}
\label{eq:stor}
\mathbf{s}({\bf{r}} ) = {\bf{r}} + v_r ({\bf{r}} ) \hat {\bf{r}},
\end{equation}
where $v_r$ is the velocity in the radial direction.

The measured clustering of galaxies will therefore be anisotropic and the additional radial signal can be used to determine the characteristic amplitude of the pair-wise distribution of the peculiar velocities at a given scale, which in turn depends on the growth rate.

Measurements are normally obtained over a small range of scales, because of simplified modeling. In this work, we use the extended analysis tested in \cite{raccanelli10} and \cite{samushia11}, which includes a more realistic description of the geometry of the system, dropping the plane-parallel approximation; this allows us to fit the observed galaxy correlation function on a larger range of scales, and therefore to be more sensitive to the cosmological parameter variations.

By imposing the conservation of the number of galaxies we can derive the Jacobian for the real- to redshift-space transformation (at the linear order):
\begin{equation}
\label{eq:jacobian}
  \delta^s({\bf{s}}) = \delta^r({\bf{r}})-
    \left( \frac{\partial v}{\partial r}+\frac{\alpha({\bf{r}})v}{r} \right),
\end{equation}
where $\delta^{s,r}$ are the observed redshift- and real-space galaxy overdensity at positions s and r, and:
\begin{equation}
\label{eq:alpha}
  \alpha({\bf{r}}) = \frac{\partial \ln r^2 \bar{N}^r({\bf{r}})}{\partial \ln r},
\end{equation}
and $\bar{N}^r({\bf{r}})$ is the expected galaxy distribution in real space.
The simplest statistic that can be constructed from the overdensity field is the correlation function $\xi(r_{12})$, defined as:
\be
\label{eq:xidef}
  \xi(r_{12}) \equiv
    \langle \delta({\bf{r}}_1) \hspace{1pt} \delta({\bf{r}}_2) \rangle \, .
\ee
In linear theory, all of the information is enclosed in the first three even coefficients of the Legendre polynomial expansion of the function $\xi$~\citep{hamilton92}:
\be
\xi(r,\mu) = \xi_0(r) L_0(\mu) + \xi_2(r) L_2(\mu) + \xi_4(r) L_4(\mu) \, ,
\ee
where $L_\ell$ are the Legendre polynomials and $\mu$ is the cosine of the angle with the line of sight ($\mu = cos(\phi)$ in Figure~\ref{fig:triangle}).

In this work we will use measurements of the momenta of the correlation function from the SDSS DR7 catalogue, using a methodology that takes care of several corrections, as we will describe in the next Sections.


\section{The Sloan Digital Sky Survey}
\label{sec:sdss}
We use data from the SDSS-II data release 7 (DR7), which obtained wide-field CCD photometry
(\citealt{gunn98}) in five passbands ($u,g,r,i,z$; e.g. \citealt{fukugita96}),
amassing nearly 10,000 square degrees of imaging data for which the object
detection is reliable down to $r \sim 22$ \citep{abazajian09}.  From these
photometric data, Luminous Red Galaxies (LRG) were targeted
\citep{eisenstein01} and spectroscopically observed, yielding a sample
of 106,341 LRGs in the redshift bin $0.16<z<0.44$.

An estimate of the statistical errors associated with the measurements is achieved through LasDamas mock catalogues, which model the clustering of Sloan Digital Sky Survey (SDSS) galaxies in the redshift span $0.16<z<0.44$. The simulations are produced by placing artificial galaxies inside dark matter halos using an HOD with parameters measured from the SDSS galaxy sample. We use the 80 ``Oriana'' catalogs that have exactly the same angular mask as the SDSS survey and subsample them to match the redshift distribution of the Luminous Red Galaxies in our SDSS DR7 data set.
Further details regarding the angular and redshift distribution of galaxies in random catalogues can be found in~\cite{samushia11}.


\subsection{Methodology}
\label{sec:method}
To perform our analysis of cosmological models we use the methodology presented in~\cite{raccanelli10} and measurements of SDSS-II data from~\cite{samushia11}.
In the Appendix we briefly revisit the main aspects of the approach we followed; this drops the distant observer approximation and includes a careful treatment of survey geometry, non-linear effects and the distribution of pair orientation.
This allows us to consider a wide range of scales (we use measurements from 30 to 120 Mpc/h).


\section{Parameterizing the Growth of Structure}
\label{sec:growth}

Measuring the matter velocity field at the locations of the galaxies gives an 
unbiased measurement of $f\sigma_{8 \rm m}$, provided that the distribution of galaxies randomly samples matter velocities, where:
\begin{equation}
\label{eq:flogd}
f = \frac{d\ln D}{d\ln a} \, ,
\end{equation}
is the logarithmic derivative of the linear growth rate, $D(a)\propto \delta_m$, with respect to the scale factor $a$ ($\delta_m$ being the fractional matter density perturbation) and $\sigma_{8 \rm m}$ quantifies the amplitude of fluctuations in the
matter density field.

\cite{linder05} proposed a gravitational growth rate formalism, which parameterises the growth factor as:
\be
\label{eq:growthlinder}
D(a) = a \mbox{ {\rm exp}} \left[ \int_0^a \left[ \Omega_m^{\gamma}(a')-1 \right] \frac{d a'}{a'} \right],
\ee
which leads to the following expression for $f$:
\be
\label{eq:fgamma}
f = \left[\Omega_m(a) \right]^\gamma \, , 
\ee
with:
\be
\Omega_m(a) = \frac{\Omega_m a^{-3}}{\sum_i \Omega_i \mbox{ {\rm exp}} \left[ 3 \int_a^1 \left[ w_i(a')+1 \right] \frac{d a'}{a'} \right]},
\ee
where the summation index goes over all the components of the Universe (i.e. dark matter, dark energy, curvature, radiation).
Within this formalism, $\gamma$ is a parameter that is different for different cosmological models: in the standard \lcgr model it is constant, $\gamma \approx 0.55$, while it is $\approx 0.68$ for the self-accelerating DGP model (see e.g. \citealt{linder05}). In some other cases, it is a function of the cosmological parameters or redshift, as we will discuss. 
It should be noted, however, that the parameterization given by Equation~\ref{eq:growthlinder} does not necessarily describe the growth rate in non-standard cosmologies.

Given that, as we will see, $\xi_\ell$ depend on $f$, measuring RSD allows us to determine $\gamma$, and hence to test different cosmological models. This also provides a good discriminant between modified gravity and dark energy models, as argued by \cite{linder05, linder07, guzzo08}.


\subsection{Degeneracy $\theta$-$\gamma$}
\label{sec:degeneracy}
In Appendix~\ref{sec:appmethod} we review the methodology used to obtain measurements of multipoles of the correlation function; our methodology involves corrections with respect to standard analyses, and in particular it drops the distant observer approximation ($\theta=0$ in Figure~\ref{fig:triangle}).
The error in the estimate of the correlation function induced by assuming $\theta=0$ can lead to a wrong estimate of the cosmological parameters measured with Redshift-Space Distortions.

\begin{figure}
\centering
\includegraphics[width=1\columnwidth]{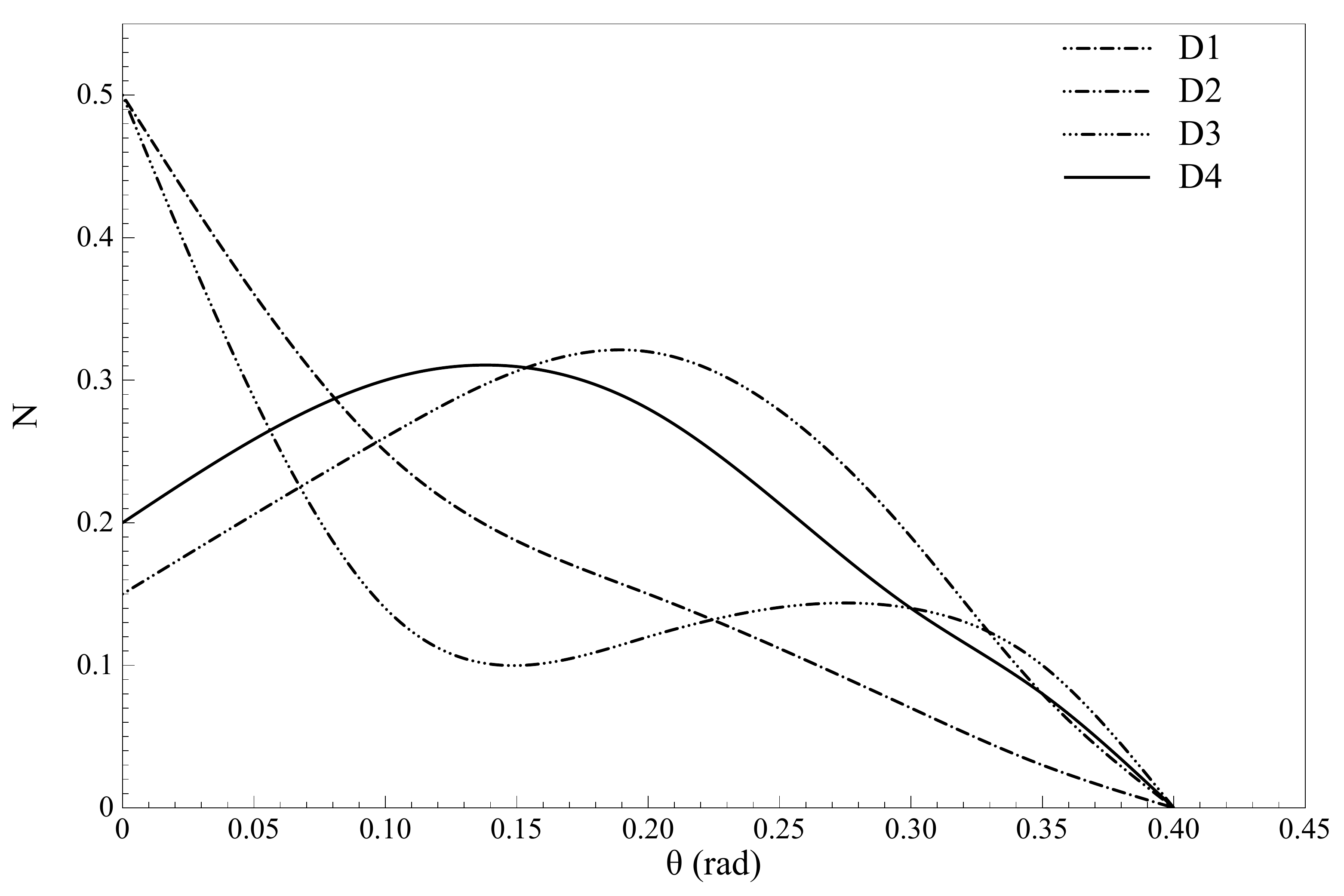}
\caption{Mock $\theta$ distributions used to illustrate effects of neglecting wide-angle corrections when measuring \gam.
\label{fig:distrib}
}
\end{figure}

\begin{figure}
\centering
\includegraphics[width=1\columnwidth]{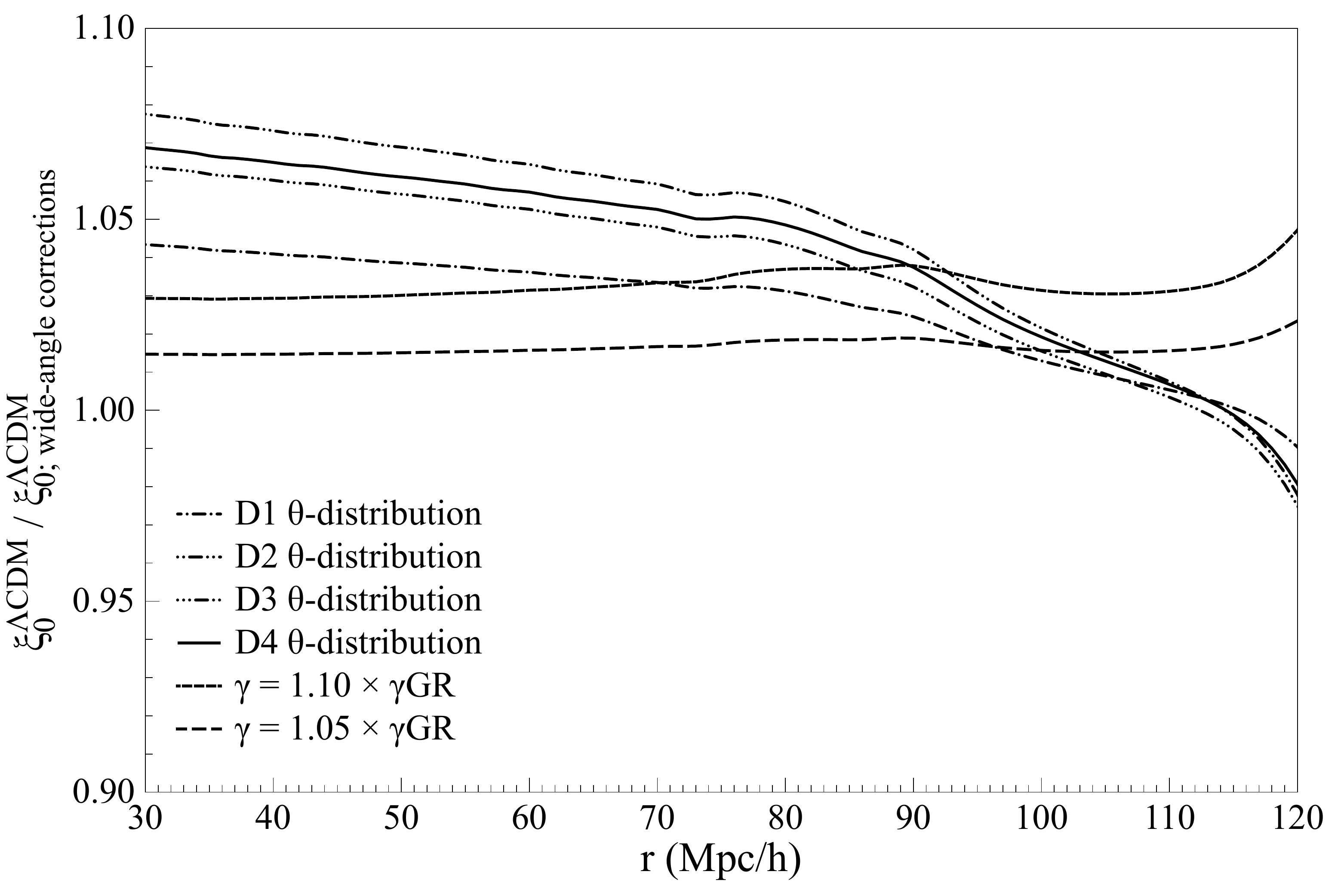}
\caption{Ratio of monopole of the correlation function in the \lcgr model in the Kaiser approximation to the monopole of the correlation function including wide-angle corrections for the mock $\theta$-distributions of Figure~\ref{fig:distrib}, compared with the effect of varying \gam in the Kaiser approximation.
\label{fig:xi0thetagammadeg}
}
\end{figure}

\begin{figure}
\centering
\includegraphics[width=1\columnwidth]{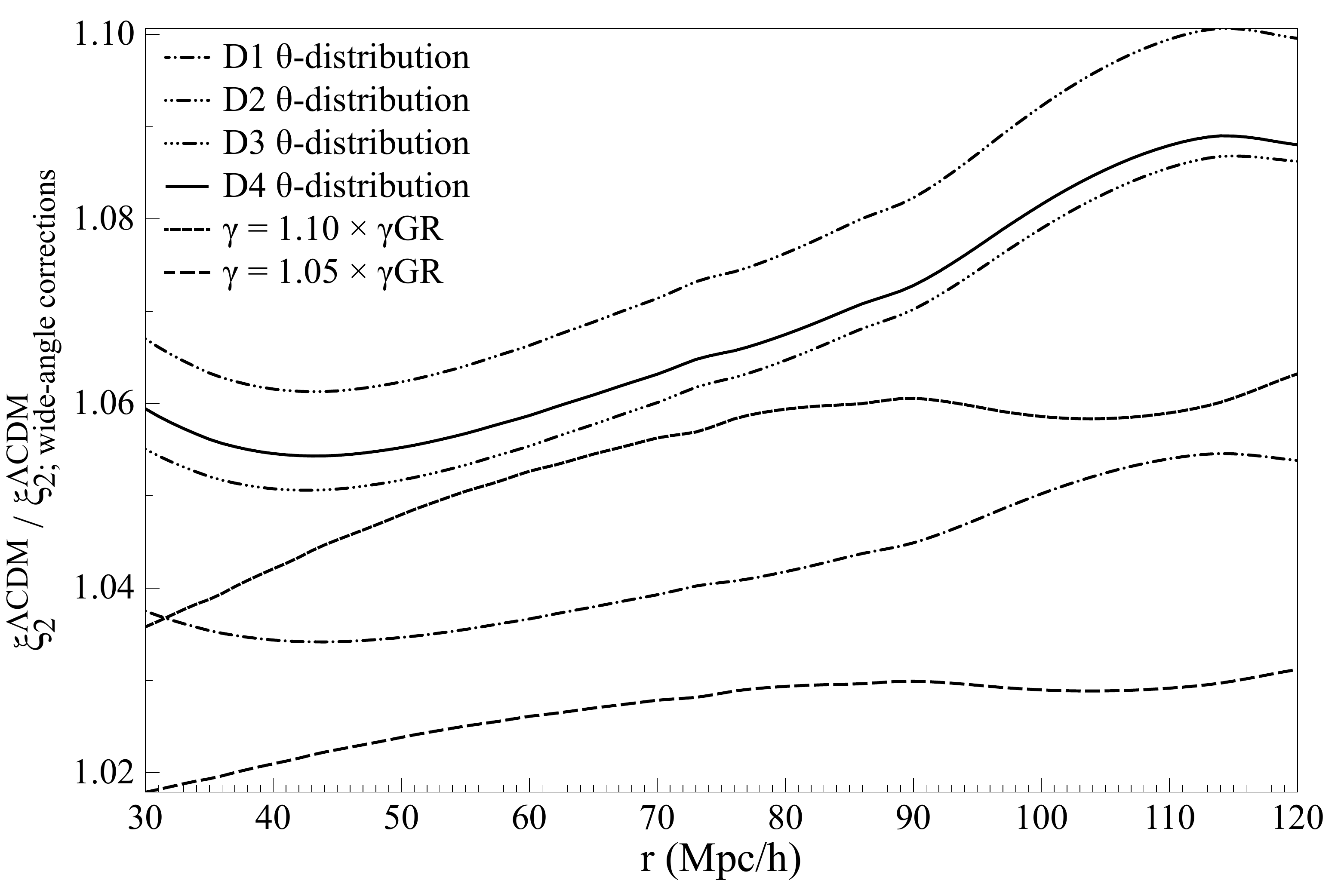}
\caption{Ratio of quadrupole of the correlation function in the \lcgr model in the Kaiser approximation to the quadrupole of the correlation function including wide-angle corrections for the mock $\theta$-distributions of Figure~\ref{fig:distrib}, compared with the effect of varying \gam in the Kaiser approximation.
\label{fig:xi2thetagammadeg}
}
\end{figure}

We investigate the systematics introduced by assuming $\theta=0$ for four different mock distributions of $\theta$ ($D1$, $D2$, $D3$, $D4$).
In Figure~\ref{fig:distrib} we plot the galaxy distribution as a function of the separation angle for each of them. The ratio between the monopole of the correlation function, $\xi_0$, derived under the Kaiser approximation and those obtained when including the wide-angle correction is shown in Figure~\ref{fig:xi0thetagammadeg}, as a function of separation scale for the standard cosmological model. The same quantity for the quadupole of the correlation function, $\xi_2$, is plotted in Figure  \ref{fig:xi2thetagammadeg}.
As expected, the largest errors are induced by the distributions D2 and D4, which have galaxy pair distributions peaked at relatively large $\theta$; however, we can note that even the D1 distribution introduces a correction of the order of $~5\%$. 
For comparison, we also add the effect that a variation of the $\gamma$ parameter would have in the Kaiser approximation. Even though the shape at large angular separation is different, the size of the variation is comparable and the wide-angle correction must be included for a proper, more exact, analysis.

The distributions used here are illustrative; a more careful analysis of the influence of wide-angle and other large-scale corrections for future Euclid-like and SKA-like surveys is left to a follow-up paper.


\section{Measurements}
\label{sec:res}
The momenta of the correlation function are sensitive to the $\gamma$ parameter through the function $f$. In this work we concentrate on two alternatives to the standard \lcgr scenario: i) the Unified Dark Matter (UDM) cosmology~\citep{Bertacca:2008uf, Bertacca:2010ct}, and ii) the normal branch DGP model including Dark Energy (nDGP) of~\cite{Schmidt09}.
These two models deviate from the standard cosmology in different ways: the UDM model assumes a single dark fluid with a clustering part, and leaves GR unmodified, while in nDGP, gravity crosses over from 4D to 5D above a cross-over scale $r_c$ (see Sections~\ref{subsec:udm} and \ref{sec:ndgp} for details). For the UDM model, we use the values of \gam of Equation~\ref{eq:gammaUDM} for a set of values of the speed of sound \cinft, while for nDGP we compute the growth solving explicitly Equation~\ref{eq:growthDGP}.

\begin{figure}
\includegraphics[width=\linewidth]{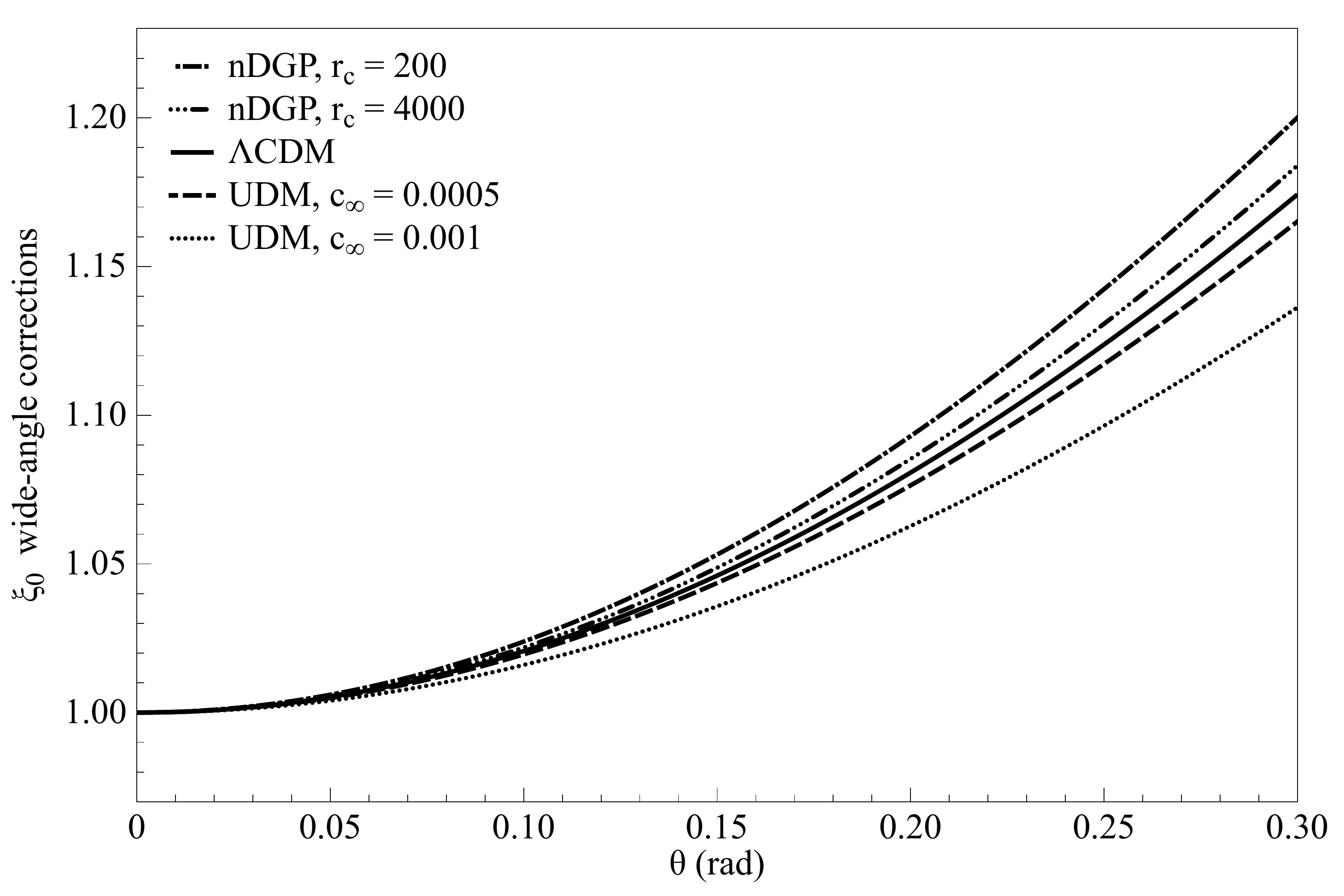}
\caption{Wide-angle corrections (i.e. $\xi_0(r,\theta)/\xi_0(r,\theta=0)$) to the monopole of the correlation function for the models considered and different values of their parameter.}
\label{fig:wa0models}
\end{figure}

\begin{figure}
\includegraphics[width=\linewidth]{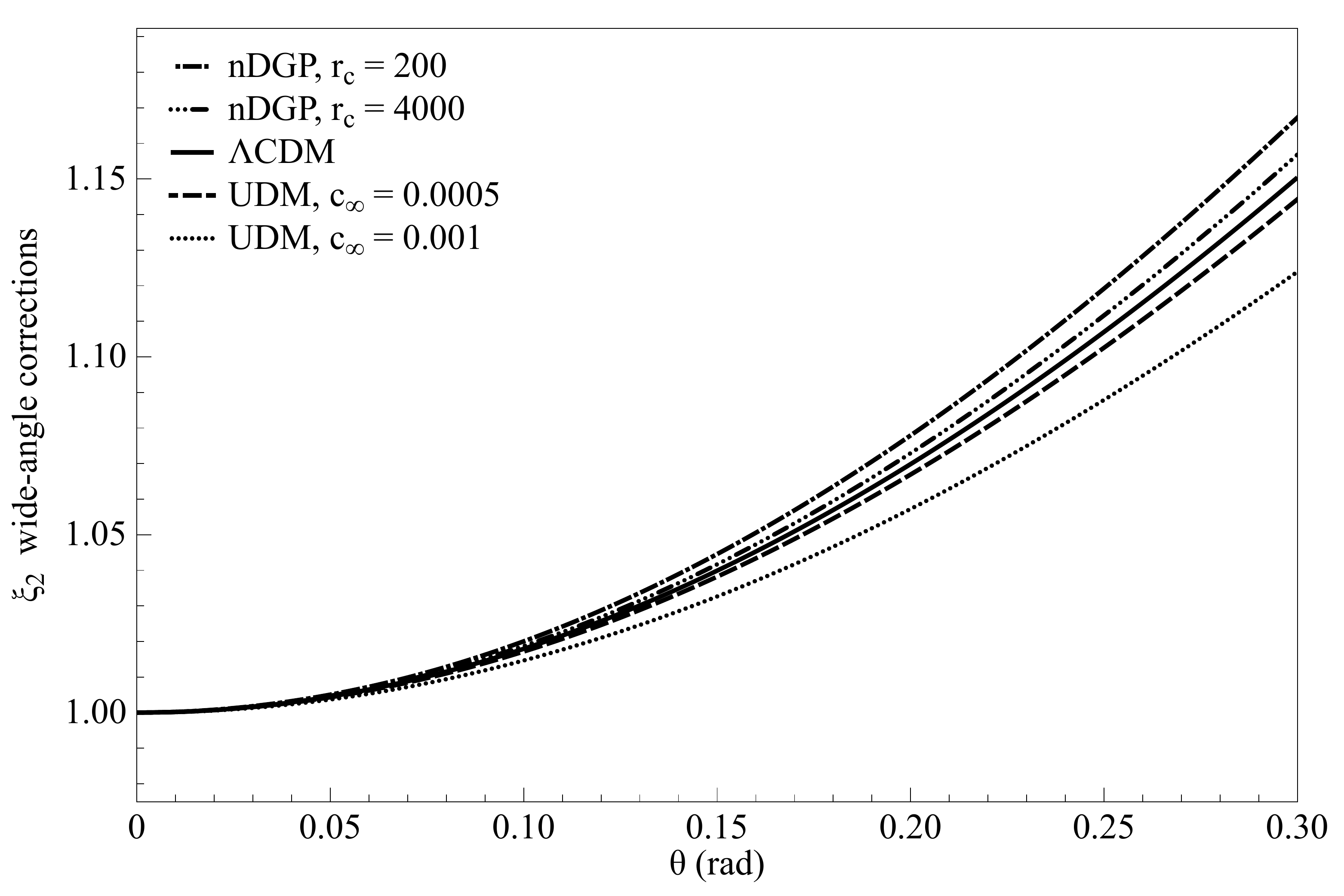}
\caption{Wide-angle corrections (i.e. $\xi_2(r,\theta)/\xi_2(r,\theta=0)$) to the quadrupole of the correlation function for the models considered and different values of their parameter.}
\label{fig:wa2models}
\end{figure}

As one can see in Equation~\ref{eq:xiwaB}, the wide-angle and mode-coupling corrections are described by a set of terms that depend on $\{r, \theta, \phi, \gamma\}$, and this means that wide-angle corrections are also different for different models of gravity. In Figure~\ref{fig:wa0models}, \ref{fig:wa2models} we show the wide-angle corrections to the monopole and quadrupole of the correlation function for the different models, compared to the \lcgr case.

\subsection{UDM}
\label{subsec:udm}
Assuming  a flat Friedmann-Lema\^{i}tre-Robertson-Walker (FLRW) background metric with scale factor $a(t)$, \cite{Bertacca:2008uf, Bertacca:2010ct}, introduced a class of  Unified Dark Matter  (UDM)  scalar field models which, by allowing a pressure equal to $-c^2 \rho_\Lambda$ on cosmological scales, reproduces the same background expansion as the $\Lambda$CDM one. In such a way, a single scalar field can be responsible for both the late time accelerated expansion for the Universe and of the growth of structure. When the energy density of radiation becomes negligible, the background evolution of the Universe is completely described by:
\begin{equation}
\label{HLcdm}
H^2(z) = H_0^2 \left[ \Omega_{\Lambda 0} + \Omega_{\rm m0}\left(1 + z\right)^{3} \right] \;,
\end{equation}
where $H$  is the Hubble parameter, $H_0=H(z=0)$ and $z$ is the redshift. $\Omega_{\Lambda 0}$ and $\Omega_{\rm m0} =1 -\Omega_{\Lambda 0} $ can be interpreted as the ``cosmological constant" and ``dark matter'' density parameters, respectively.

The density contrast of the clustering component is $\delta_{\rm DM}\equiv \delta \rho/\rho_{\rm DM }$, where $\rho_{\rm DM}=\rho -\rho_\Lambda$ is  the only component of the scalar field density that clusters. In these models one of the most relevant parameters is the sound speed of the perturbations, which defines a typical sound-horizon (Jeans length) scale above which growth of structure is possible~(\citealt{Bertacca:2007cv}). Following \cite{bertacca11}, for scales smaller than the cosmological horizon and $z < z_{\rm rec}$, we have that:
\begin{equation}
\delta_{\rm {DM}}\left[k;\eta(z)\right]  = T_{\rm UDM}\left[k;\eta(z)\right] \delta_m \left[k;\eta(z)\right] \;,
\end{equation}
where $\delta_m$ is the matter density perturbation in the standard $\Lambda$CDM model, $\eta$ is the conformal time and $T_{\rm UDM}(k;\eta)$ is the transfer function for the UDM model:
\begin{equation}
\label{eq:fittfunc}
T_{\rm UDM}(k;\eta) = j_0[ \mathcal{A}(\eta) k] \;,
\end{equation}
\begin{equation}\label{Aformula}
\mathcal{A}(\eta) = \int^{\eta}c_{\rm s}(\eta') \; \ud \eta' ,
\end{equation}
\begin{equation}
\label{eq:cs2}
{c^2_{\rm s}}(a)=\frac{{\Omega_{\Lambda 0} c^2_\infty}}{\Omega_{\Lambda 0}+(1-{c^2_\infty})\Omega_{\rm m 0}a^{-3}}\;.
\end{equation}
We then define the parameter $c_\infty$ as the value of the sound speed when $a\rightarrow\infty$.

\begin{figure}
\includegraphics[width=\linewidth]{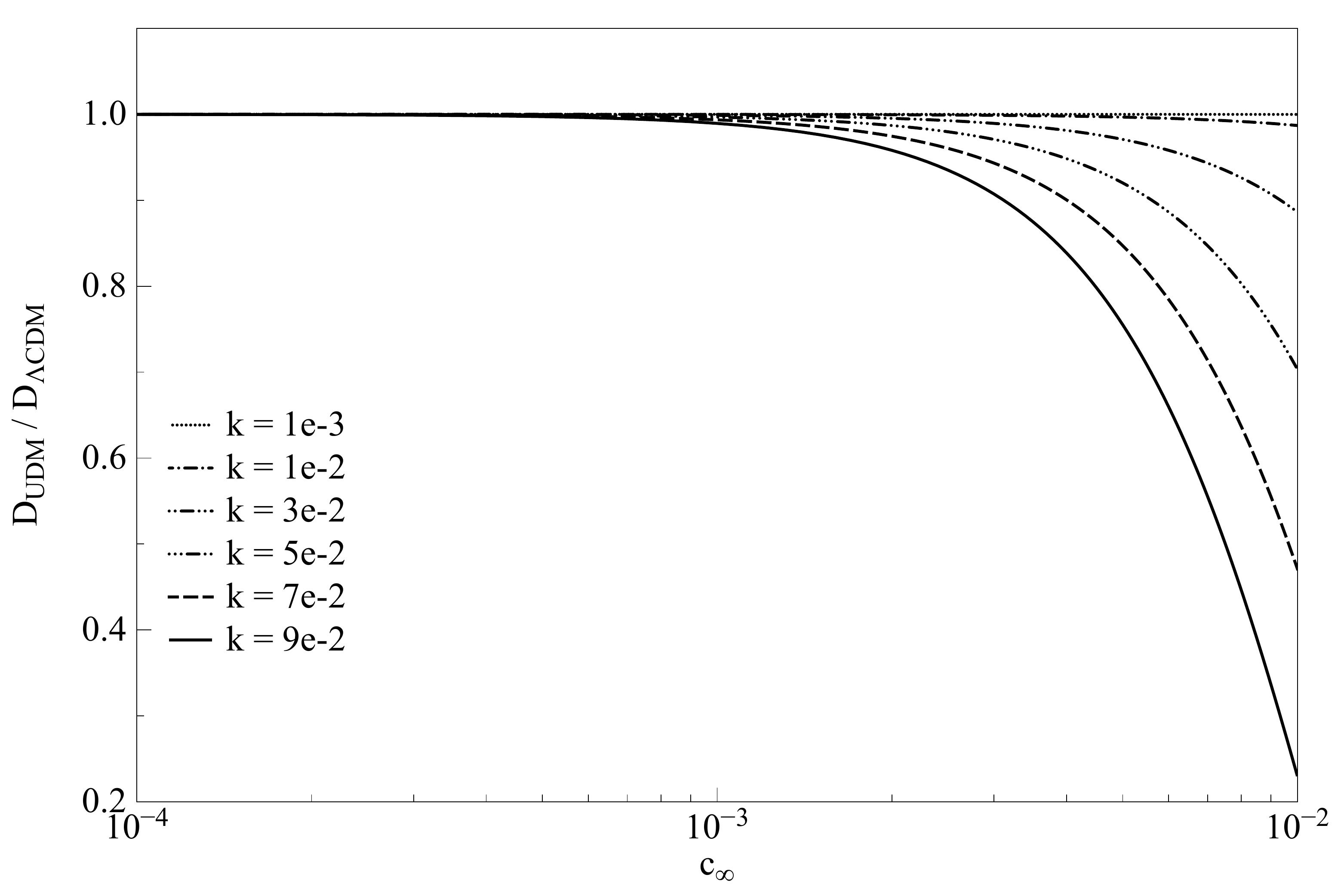}
\caption{Deviation of the growth factor for the UDM model, as a function of the speed of sound \cinft, for different values of $k$.}
\label{fig:growthudm}
\end{figure}

In this model, the growth factor is computed correcting the \lc one with the transfer function of Equation~\ref{eq:fittfunc}; its deviation from the standard model case, as a function of \cinft for different values of $k$, is plotted in Figure~\ref{fig:growthudm}.

We refer to \cite{Bertacca:2008uf, Bertacca:2010ct, bertacca11} for details of the UDM models considered (see also \citealt{Camera:2009uz, Camera:2010wm, Piattella:2011fv, Camera:2012sf}).

\begin{figure}
\includegraphics[width=\linewidth]{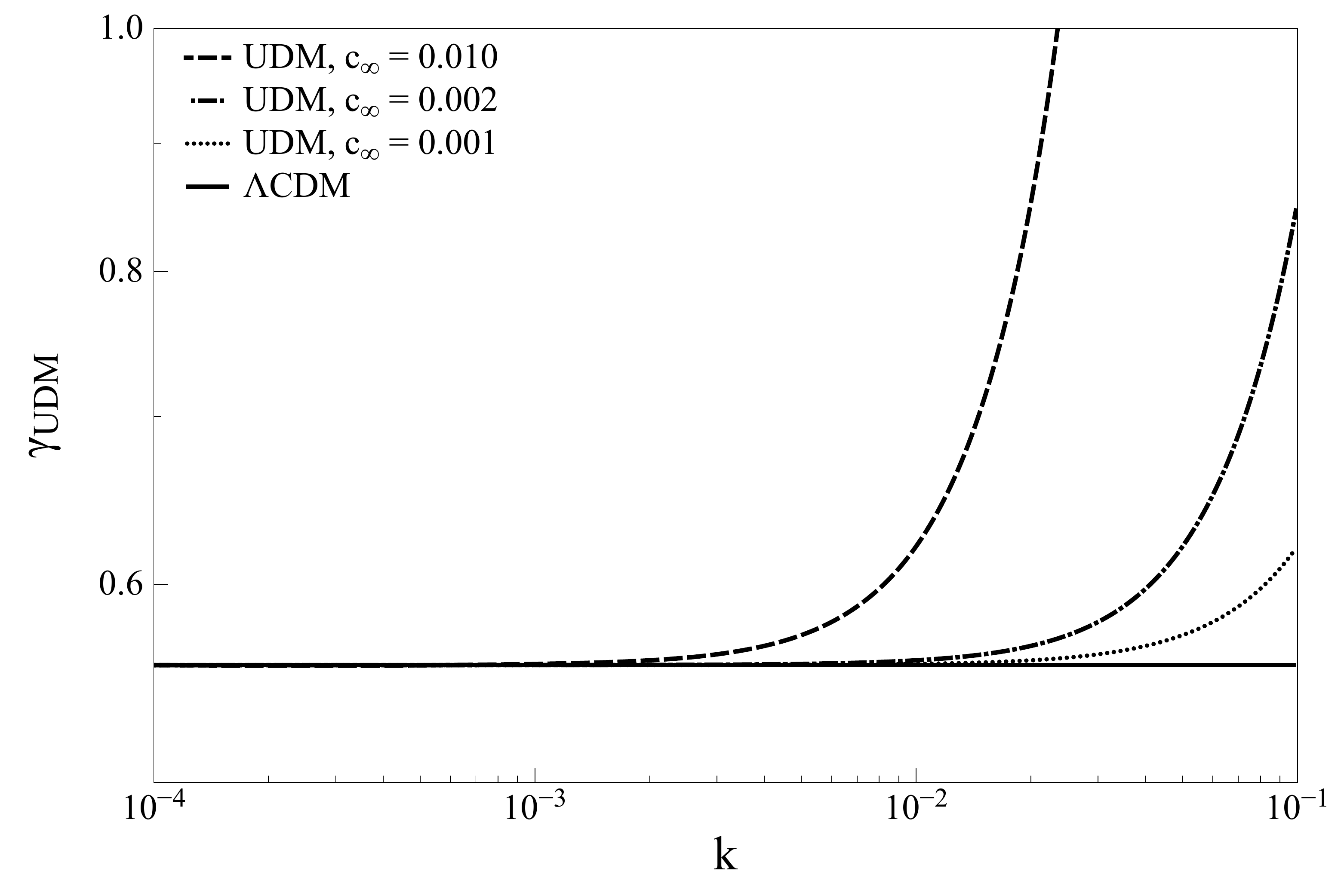}
\caption{$\gamma_{\rm UDM}$ as a function of $k$, for different values of $c_{\infty}$ (at $z=0.15$), compared with the \lcgr case.}
\label{fig:gammaUDM_z015_cinfi}
\end{figure}

Here we report a new result, which is particularly relevant for the present analysis: the parameterization of the growth rate in UDM models, which turns out to be:
\begin{equation}
\label{eq:gammaUDM}
\gamma_{\rm UDM} (a,k,c_{\infty}) =\frac{\ln \left[\frac{\ud \ln T_{\rm UDM}(a,k,c_{\infty})}{\ud \ln a}+f_{\rm {\Lambda CDM}}(a)\right]}{\ln \Omega_m(a)}\; ;
\end{equation}
as one can see, $\gamma_{\rm UDM}$ depends on $(a,k,c_{\infty})$, and in Figure~\ref{fig:gammaUDM_z015_cinfi} is shown its value as a function of them.

Equation~\ref{eq:gammaUDM} gives the value of $\gamma$ for UDM cosmologies; it depends on redshift, scale and speed  of sound. Figure \ref{fig:gammaUDM_z015_cinfi} shows that $\gamma_{\rm UDM} \geq \gamma_{\rm {\Lambda CDM}}$. Actually this is easy to understand: one of the features of the UDM models is that, under the Jeans length, the density contrast decreases while oscillating in time (see, e.g., \citealt{Bertacca:2007cv, bertacca11}). This means that, compared to the $\Lambda$CDM model, there is a further suppression in the growth of structures.  
Notice also that there are some values of $c_{\infty}$ that give non-physical results when one uses the parameterization of Equation~\ref{eq:gammaUDM}; however, even using Equation~\ref{eq:flogd}, the growth rate for these values becomes highly oscillatory, so we will consider them ruled out.
For this reason, given that we will limit our analysis to linear scales, and accounting also for existing limits on UDM models (e.g.~\citealt{bertacca11}) we will consider values of $c_{\infty} \lesssim 0.002$. 

\begin{figure}
\includegraphics[width=\linewidth]{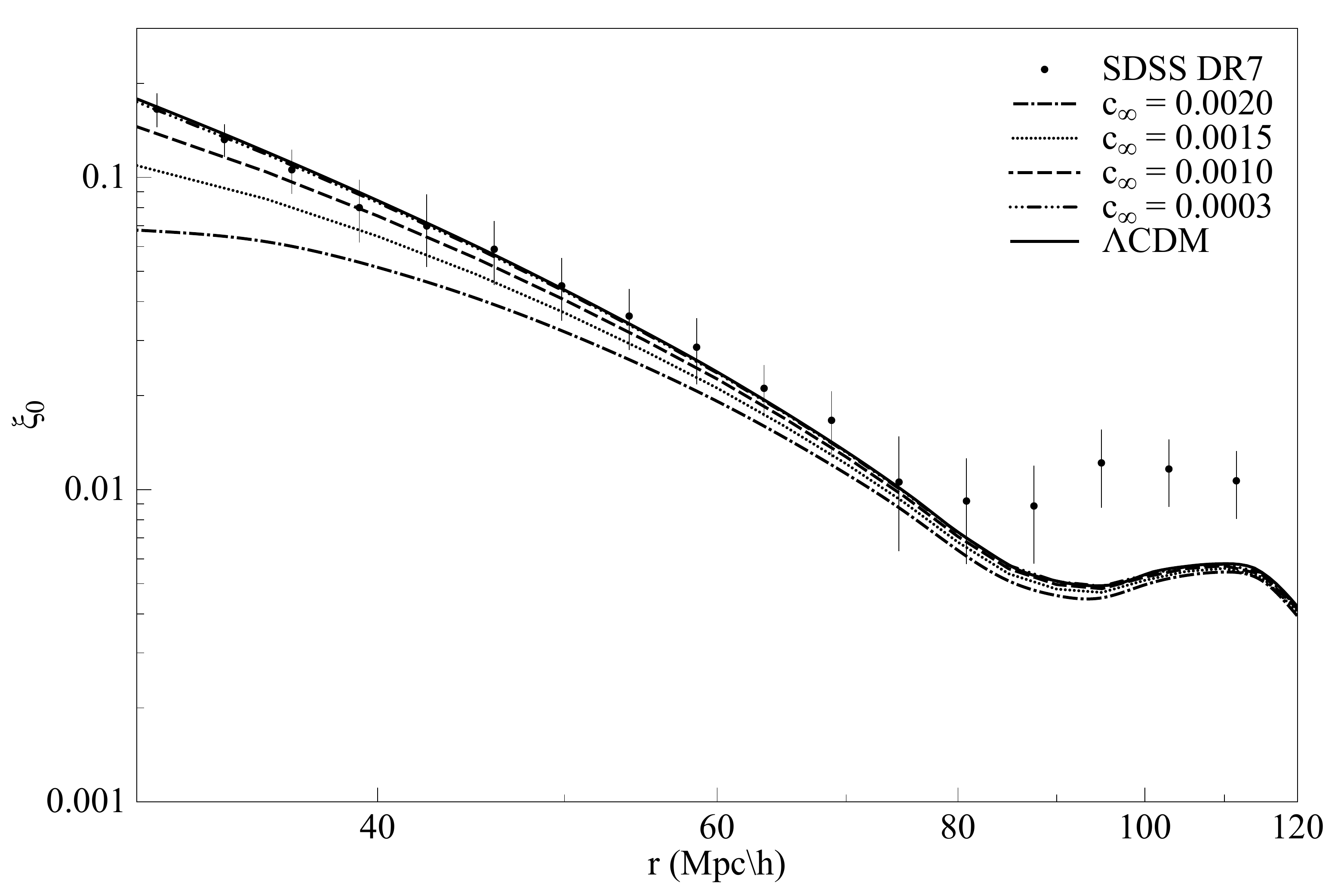}
\caption{$\xi_0$ measured from SDSS DR7 and theoretical predictions for \lc and UDM with different values of the speed of sound.}
\label{fig:xi0UDM}
\end{figure}

\begin{figure}
\includegraphics[width=\linewidth]{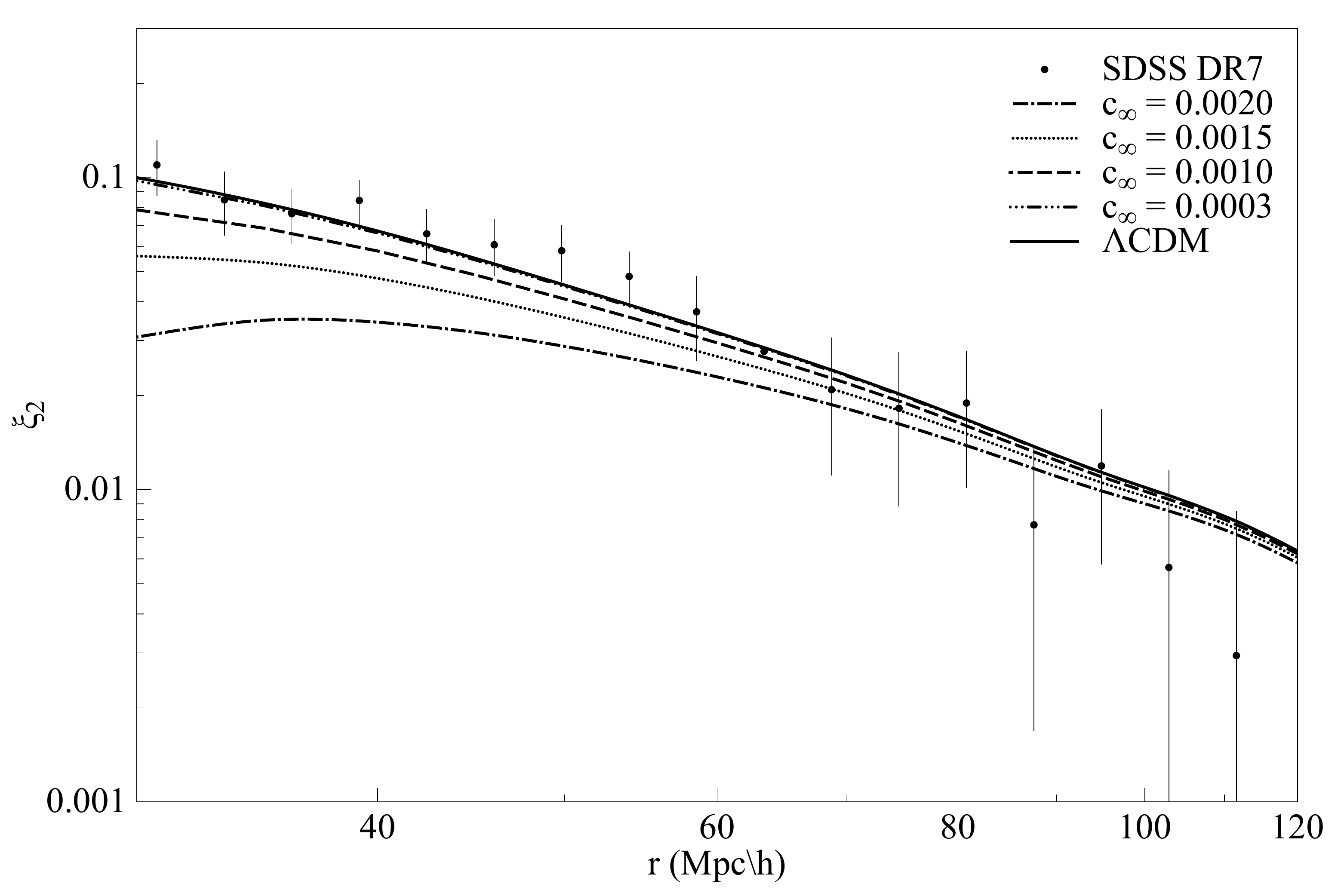}
\caption{$\xi_2$ measured from SDSS DR7 and theoretical predictions for \lc and UDM with different values of the speed of sound.}
\label{fig:xi2UDM}
\end{figure}

In Figures~\ref{fig:xi0UDM},\ref{fig:xi2UDM} we show the comparison of measurements of the zeroth and second order momenta of the correlation function measured from SDSS DR7 with the theoretical predictions for UDM models, for several values of the speed of sound parameter \cinft.

\subsection{DGP}
\label{sec:ndgp}
In the Dvali-Gabadadze-Porrati (DGP, \citealp{DGP}) model, all matter
and radiation are confined to a four-dimensional brane in five-dimensional
Minkowski space.
Gravity, while restricted to the brane on small scales,
propagates into the extra dimension above the cross-over scale $r_c$. 
This scenario admits an FRW cosmology on the brane \citep{Deffayet01},
where the Friedmann equation is modified to:
\be
H^2 \pm \frac{H}{r_c} = \frac{8\pi G}{3} \left[\bar\rho_m + \rho_{\rm DE}\right].
\ee
The sign on the left-hand-side depends on the choice of embedding of the
brane. The negative sign (\emph{accelerating branch}) leads to an
accelerated expansion of the Universe at late times without a cosmological
constant or Dark Energy \citep{Deffayet01}, i.e. $\rho_{\rm DE}=0$ 
and $r_c \sim H_0^{-1}$, while the positive sign
(\emph{normal branch}) does not yield acceleration by itself.  The simplest 
self-accelerating model is in conflict with observations of the CMB
and Supernovae (e.g., \citealt{LombriserEtal}), and also has theoretical
issues \citep{LutyEtal,NicRat,GregoryEtal}. 

Here we consider a normal-branch DGP model including a Dark Energy
component $\rho_{\rm DE}$ to yield an accelerated expansion. 
Specifically, we use the model of \cite{Schmidt09} where the equation of state of the Dark Energy
is tailored to yield an expansion history identical to $\Lambda$CDM
for all $r_c$.  This model is not ruled out by expansion history probes
or theoretical issues.  Furthermore, it can serve as a toy model for
more recent scenarios \citep{deRham,AfshordiEtal}.  

On scales much smaller than the horizon and smaller than $r_c$, 
but yet large enough so that linear perturbation theory applies,
the growth of perturbations during matter domination in these models
is governed by \citep{KoyamaMaartens}:
\begin{equation}
\ddot\delta + 2 H \dot\delta = \frac32 \Omega_{\rm m}(a) a^2 H^2 
\left(1+\frac1{3\beta(a)}\right) \delta,
\label{eq:growthDGP}
\end{equation}
where dots denote derivatives with respect to time, and
\begin{align}
\beta =\:& 1 \pm 2Hr_{\rm c} \left(1 + \frac{a H'}{3H} \right).
\end{align}
Here the positive (negative) sign holds for the normal (self-accelerating)
branch, respectively.  Note that in the latter case, $\beta < 0$ and hence
$G_{\rm eff} < G$, i.e. the growth of structure is slowed down with respect
to GR. The opposite is the case for the normal branch
($\beta > 0$).  
In the case of the self-accelerating (sDGP) model with $\rho_{\rm DE}=0$, 
\cite{linder05} showed
that the growth rate can be well described by the parameterization
$f_{\rm sDGP}(a) = \Omega_m(a)^{\gamma_{\rm sDGP}}$, with $\gamma_{\rm sDGP} = 0.68$. 
For the normal branch model with Dark Energy (nDGP), this parametrization does not provide a good fit, and we use a numerical integration of Equation~\ref{eq:growthDGP} to derive the growth rate. The change in the growth with respect to $\Lambda$CDM is shown in Figure 10 as a function of $r_c$.

\begin{figure}
\includegraphics[width=\linewidth]{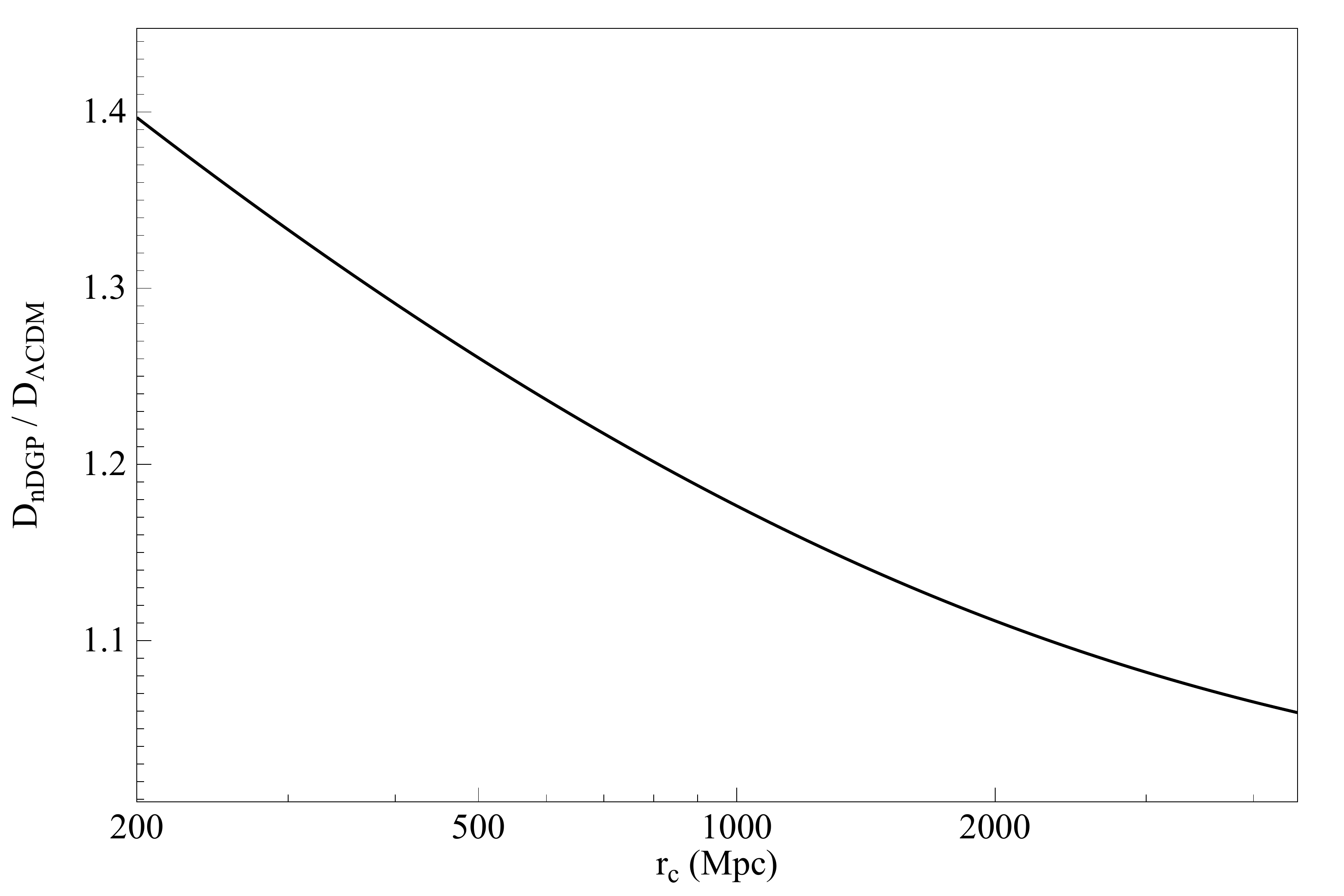}
\caption{Deviation of the growth factor for the nDGP model, as a function of the cross-over scale $r_c$ (see text for details).}
\label{fig:growthndgp}
\end{figure}

\begin{figure}
\includegraphics[width=\linewidth]{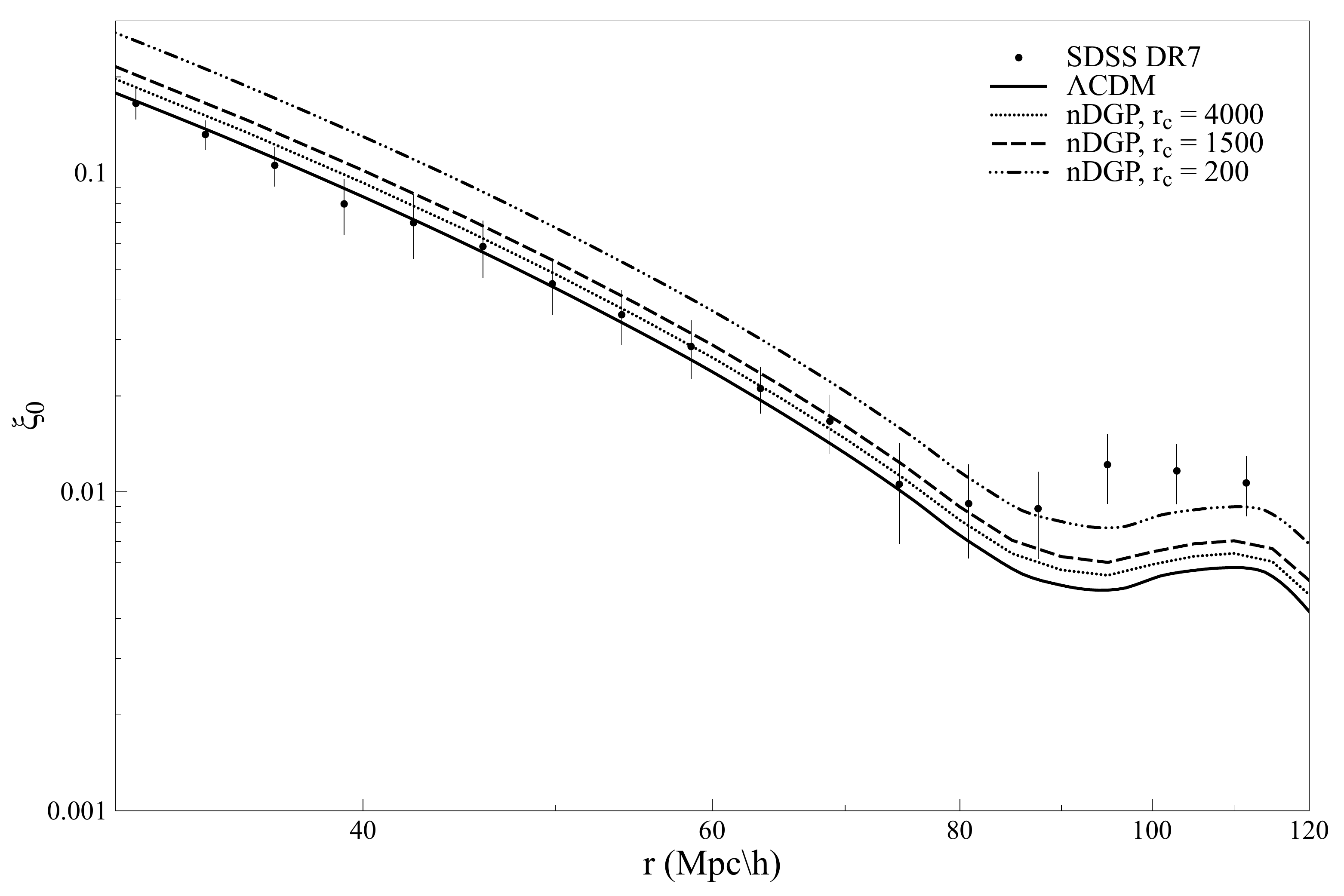}
\caption{$\xi_0$ measured from SDSS DR7 and theoretical predictions for \lc and nDGP, for some values of the cross-over scale $r_c$.}
\label{fig:xi0nDGP}
\end{figure}

\begin{figure}
\includegraphics[width=\linewidth]{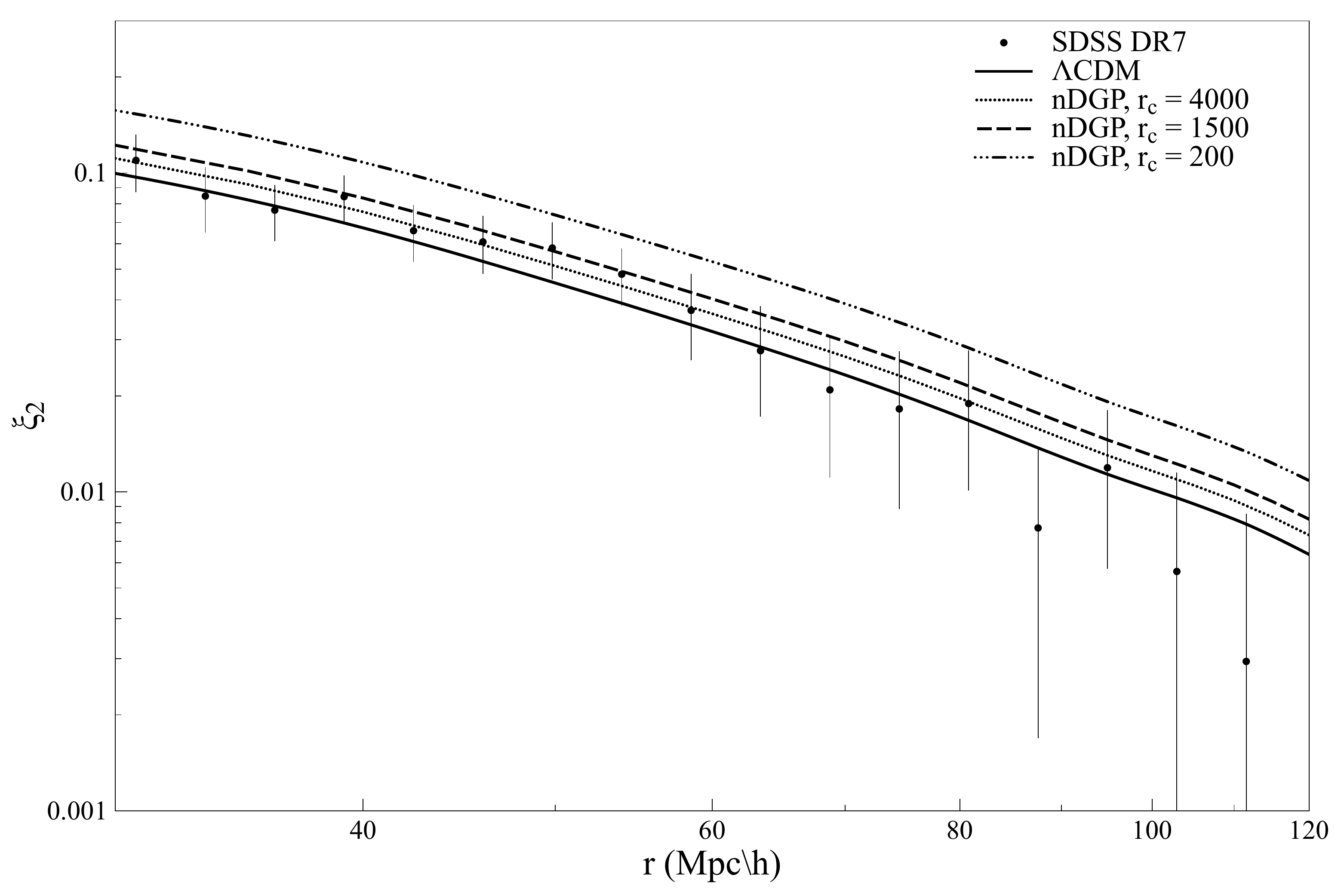}
\caption{$\xi_2$ measured from SDSS DR7 and theoretical predictions for \lc and nDGP, for some values of the cross-over scale $r_c$.}
\label{fig:xi2nDGP}
\end{figure}

In Figures~\ref{fig:xi0nDGP}, \ref{fig:xi2nDGP} we show the comparison of the zeroth and second order momenta of the correlation function measured from SDSS DR7 with the theoretical prediction for the nDGP model, for some values of the cross-over scale $r_c$.

\section{Results}
\label{sec:results}
We compute the likelihood of the \cinft\ and $1/r_c$ parameter for UDM and nDGP model respectively, conditioned to the other parameters fixed to the WMAP 7-year best-fit \lc\ model ones.
We assume a Gaussian likelihood with covariance matrix described by Equation~\ref{eq:cstat}.
This choice has the advantage of having the \lc\ model as an asymptotic limit, when the extra parameter tends to zero.

We evaluated a joint $\mathcal{L}$ for the zeroth and second momenta of the correlation function, focusing on two cases: (i) $r_{\rm max}$ = 80 Mpc/h, where the data start to deviate from the mock for both $\xi_0$ and $\xi_2$ (see Figure~\ref{fig:xi0meascomp}, \ref{fig:xi2meascomp}), and (ii) $r_{\rm max}$ = 120 Mpc/h, in order to investigate larger scales and where the data are still in reasonable agreement with the mocks. For larger scales, \citet{ross11} suggested (when considering a different sample) that the excess power can be due to systematic errors, so, to be conservative, we will not fit these scales.

The bias is poorly understood: in order to take into account its theoretical uncertainty, we leave it as a free parameter and marginalize over it. Since additional information on the bias can be independently inferred from other probes, as for example lensing measurements, we also derive the constraints on cosmological parameters which can be obtained when fixing the bias at its best-fit value assuming a \lc cosmology.

\begin{figure}
\includegraphics[width=\linewidth]{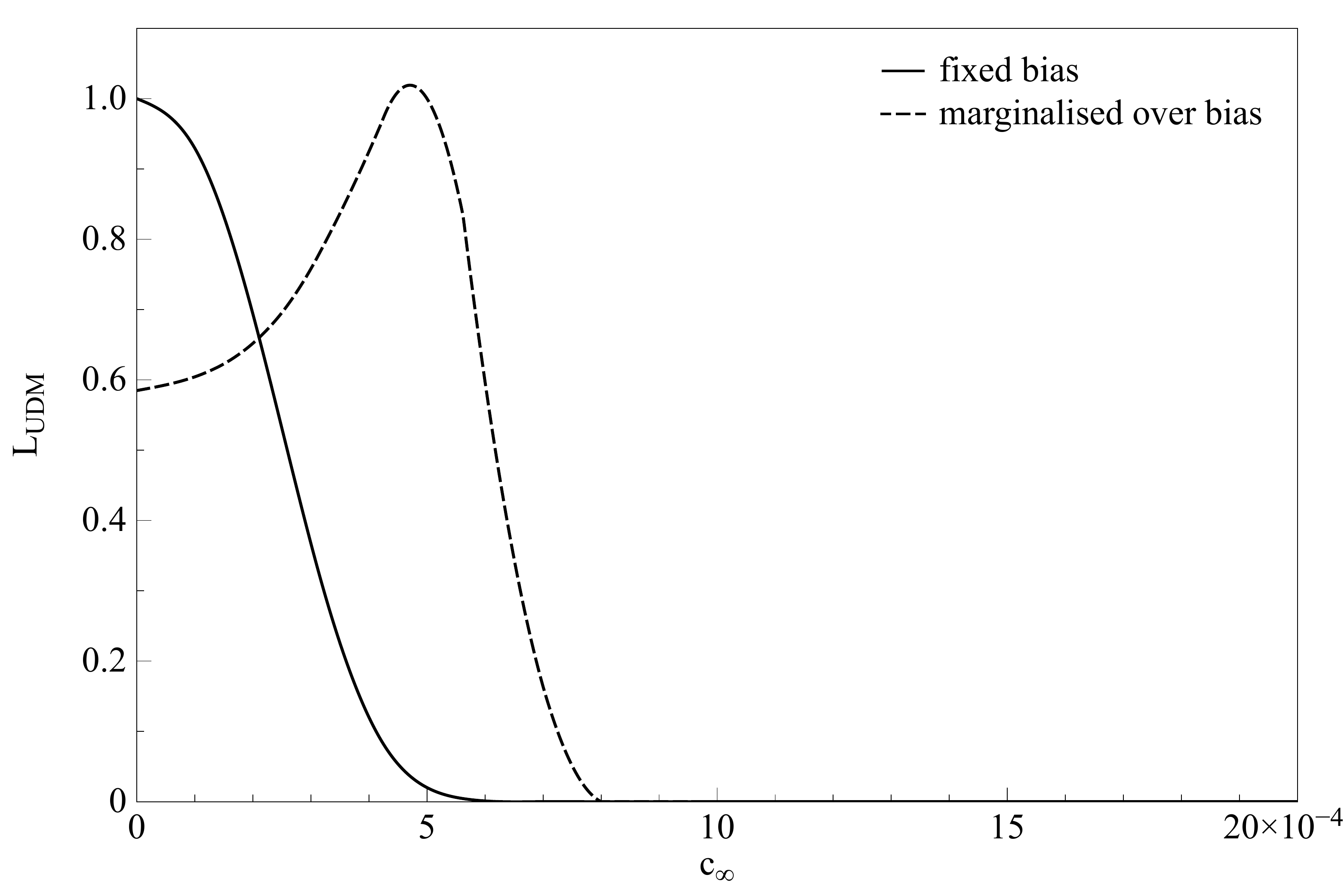}
\caption{Likelihood for the UDM model, as function of the speed of sound \cinft, when assuming knowledge of the bias (solid line), and after marginalizing over it (dashed line).}
\label{fig:L_xi02_UDM}
\end{figure}

\begin{figure}
\includegraphics[width=\linewidth]{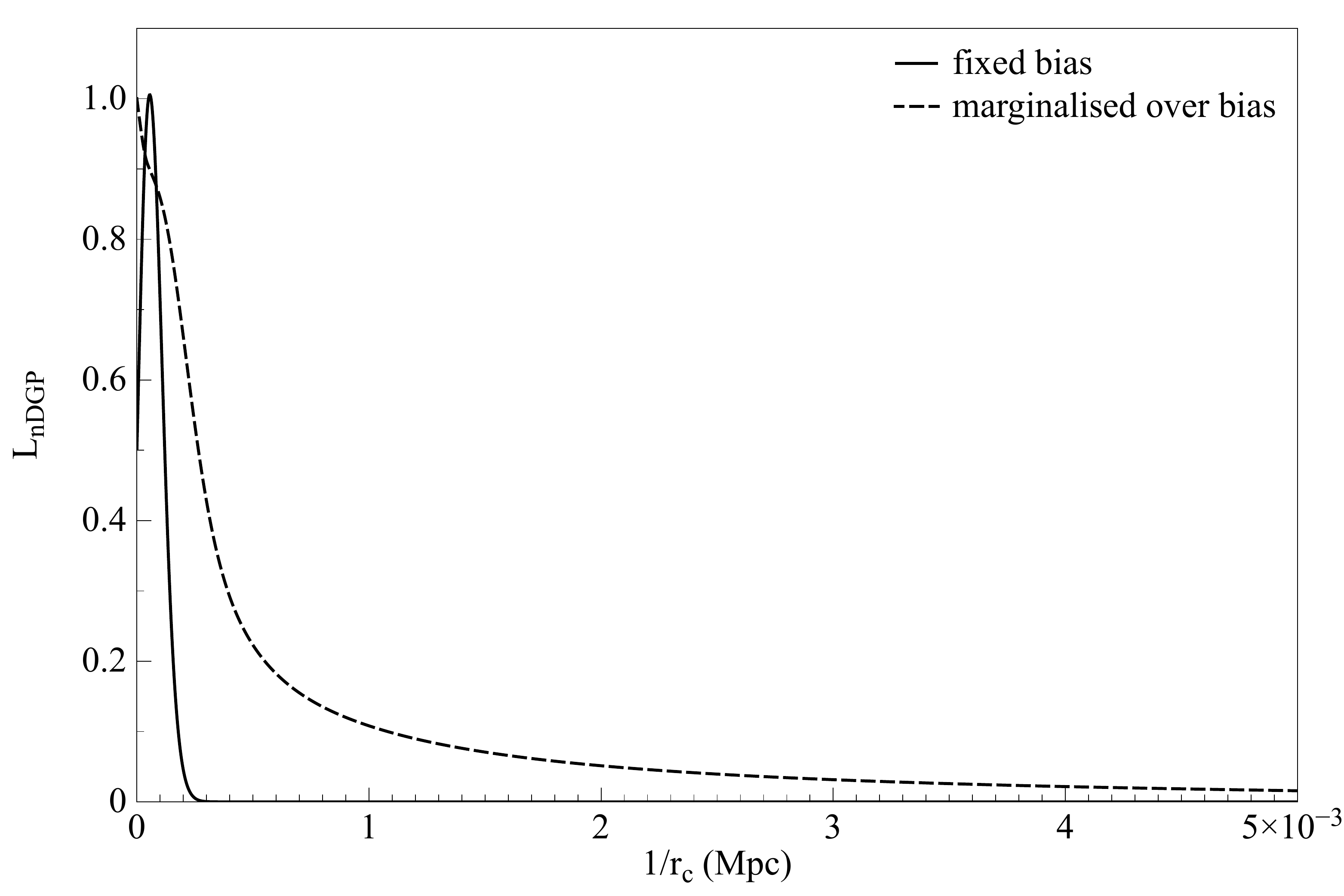}
\caption{Likelihood for the nDGP model, as function of the inverse of the cross-over scale $1/r_c$, when assuming knowledge of the bias (solid line), and after marginalizing over it (dashed line).}
\label{fig:L_xi02_nDGP}
\end{figure}

In Figures~\ref{fig:L_xi02_UDM} and \ref{fig:L_xi02_nDGP} we show the likelihood, as a function of \cinft (UDM) and $1/r_c$ (nDGP), in the $r_{\rm max}$ = 120 Mpc/h case, assuming the knowledge of the bias (solid lines), and after marginalizing over it (dashed lines).

In Table~\ref{tab:const} we report the best fit and the constraints on \cinft and $r_c$ at different confidence levels, for the various cases considered.

The bias plays a very important role: this was expected, since the shape of the correlation function is a weak function of the additional parameter in both classes of models, whereas the amplitude variation is large (in UDM the shape variation is more relevant, and increasingly important for larger values of \cinft).

In the UDM case, assuming the knowledge of the bias results in a best fit value of $\cinf=0$, which corresponds to the \lc\ model, while, after marginalizing over the bias, the best fit is $\cinf=5\times10^{-4}$. From the Table \ref{tab:const} emerges that the constraints on the parameters are not sensitive to the maximum value of the distance between galaxy pairs. This is not surprising, as the deviation of the model from \lc\ is larger for smaller scales (this can also be seen from Figures~\ref{fig:xi0UDM}, \ref{fig:xi2UDM}).
It is interesting to note that, after bias marginalisation, the best fit does not correspond to \lc, meaning that the shape of the measured correlation functions is not fitted by \lc; it would be interesting to perform a future analysis at different redshift or with other data sets (e.g. BOSS), to see how the peak of the likelihood will be modified.
Our final constraint at 95\% confidence level, after marginalizing over the bias, is \cinft $\le 6.1\times10^{-4}$, almost two orders of magnitude better than previous constraints~\citep{bertacca11}.

When considering the nDGP scenario, the effect of marginalizing over the bias is even more dramatic: the constraints on $r_c$ worsen by a factor of $\sim$20 (see Figure~\ref{fig:L_xi02_nDGP} and Table \ref{tab:const}). This occurs because the deviation from \lcgr of this model is scale independent (below the cross-over scale), hence degenerate with the bias.
After bias marginalisation, the best fit corresponds to the \lcgr model.  
However, for this model there are no published constraints in the literature, so our analysis presents a first result on that. Note that a cross-over scale of $r_c \sim 300$~Mpc implies strong modifications to gravity on larger scales, and the integrated Sachs-Wolfe (ISW) effect in the cosmic microwave background is likely able to constrain such values as well.

\begin{table}
\begin{center}
\begin{tabular}{ |p{1.55cm}|p{1.45cm}|p{1.45cm}|p{1.45cm}|p{1.45cm}| }
	  \hline
	  \textbf{Model} & \textbf{1-$\sigma$} & \textbf{2-$\sigma$} & \textbf{3-$\sigma$} & \textbf{Best Fit} \\
	  \hline
	  UDM[80], fixed bias & \cinft $\le$ 2.1e-4 & \cinft $\le$ 3.8e-4 & \cinft $\le$ 4.7e-4 & \cinft = 0 (\lc) \\
	  \hline
	  UDM[80] marginalized& \cinft $\le$ 5.8e-4 & \cinft $\le$ 6.3e-4 & \cinft $\le$ 6.9e-4 & \cinft = 5.0e-4\\
	  \hline
	  UDM[120], fixed bias & \cinft $\le$ 1.9e-4 & \cinft $\le$ 3.5e-4 & \cinft $\le$ 4.3e-4 & \cinft = 0 (\lc)4 \\
	  \hline
	  UDM[120] marginalized& \cinft $\le$ 5.8e-4 & \cinft $\le$ 6.1e-4 & \cinft $\le$ 6.9e-4 & \cinft = 5.0e-4 \\
	  \hline
	  nDGP[80], fixed bias & $r_c \ge$ 12580 (Mpc) & $r_c \ge$ 7040 (Mpc) & $r_c \ge$ 5500 (Mpc) & $r_c$ = 25000 (Mpc) \\
	  \hline
	  nDGP[80] marginalized& $r_c \ge$ 660 (Mpc) & $r_c \ge$ 290 (Mpc) & $r_c \ge$ 255 (Mpc) & $r_c$ = $\infty$ (\lc)\\
	  \hline
	  nDGP[120], fixed bias & $r_c \ge$ 9803 (Mpc) & $r_c \ge$ 6480 (Mpc) & $r_c \ge$ 5050 (Mpc) & $r_c$ = 20000 (Mpc) \\
	  \hline
	  nDGP[120] marginalized& $r_c \ge$ 1237 (Mpc) & $r_c \ge$ 340 (Mpc) & $r_c \ge$ 270 (Mpc) & $r_c$ = $\infty$ (\lc)\\
	  \hline
	\end{tabular}
\caption{Constraints and best fits for models considered, when knowledge of the bias is assumed and after marginalizing over it as a free parameter, for the $r_{\rm max}$=80 and 120 Mpc/h cases.}
\label{tab:const}
\end{center}
\end{table}


\section{Conclusions}
\label{sec:disc}
In this work, we showed that we can use Redshift-Space Distortions to test cosmological models, by measuring the monopole and quadrupole of the correlation function of galaxies, $\{ \xi_0, \xi_2 \}$.
The methodology and robust measurement of the correlation function presented in~\cite{raccanelli10, samushia11}, which includes a careful treatment of corrections due to the geometry of the system, allows us to use those multipoles in a wider range of scales w.r.t. most standard analyses, and so have better constraints on the models tested.
As shown in~\cite{samushia11}, most of the approximations assumed in the Kaiser analysis need to be dropped to have precise measurements of the correlation function, and so of the growth rate, whose deviation from the \lcgr expected value would imply the need of a new cosmological model.

We explored the Unified Dark Matter~\citep{Bertacca:2008uf, Bertacca:2010ct} and the normal branch DGP~\citep{Schmidt09} models, that present deviations from the \lcgr scenario. Both classes of models are parameterised by one additional number: the speed of sound \cinft and the cross-over scale $r_c$, for UDM and nDGP respectively. The value of these parameters affects both the growth rate parameter $\gamma$ and the wide-angle corrections. Moreover, UDM models are characterised by a growth rate parameter, $\gamma$, that depends on $\{k, z, \cinf\}$. After deriving its analytic expression, we used it to compute the predicted momenta of the correlation function.

We then compared observations of LRGs from SDSS-II DR7 with theoretical predictions from the two cosmologies. This analysis allowed us to tighten the constraints on the speed of sound of UDM models to \cinft $\le 6.1\times10^{-4}$, and to put, for the first time, a lower bound on the cross-over scale for the nDGP model of 340 Mpc, both at 95\% confidence level.
It is worth noting that the results would largely benefit from a better knowledge of the bias that could be obtained combining information from other cosmological probes, as e.g. gravitational lensing.

We showed the potential of this methodology for constraining alternative cosmological models, in particular when future surveys will provide access to a wider range of scales, hence allowing a tomographic analysis, which will be essential to break degeneracy between cosmological parameters.


\section*{Acknowledgments}
Part of the research described in this paper was carried out at the  
Jet Propulsion Laboratory, California Institute of Technology, under a  
contract with the National Aeronautics and Space Administration.
AR would like to thank for the hospitality the Institute of Cosmology and Gravitation at the University of Portsmouth, 
where part of this work was carried out.
FS is supported by the Gordon and Betty Moore Foundation at Caltech.
WJP is grateful for support from the European Research Council and STFC. 
LS acknowledges support from European Research Council, 
Georgian National Science Foundation grant ST08/4-442 and SNSF (SCOPES grant No 128040).

\appendix


\section{Methodology}
\label{sec:appmethod}

\subsection{The wide-angle corrections}
\label{sec:warsd}
Most previous RSD analyses have used the simple plane-parallel approximation given by the Kaiser formula. In this case, a Fourier mode $\hat{\delta}^s (\textbf{k})$ in redshift space is simply equal to the unredshifted mode $\hat{\delta} (\textbf{k})$ amplified by a factor $1 + \beta \mu^2_{\textbf{k}}$:
\begin{equation}
\label{eq:dsk}
\hat{\delta}^s (\textbf{k}) = (1 + \beta \mu^2_{\textbf{k}})\hat{\delta} (\textbf{k}),
\end{equation}
where $\beta = f/b$, with $b$ being the bias. 
This correction arises from the Jacobian of Equation~\ref{eq:jacobian}, when the $(1+\frac{v}{r})^{2}$ term is neglected.

The wide-angle linear redshift-space correlation function and power spectrum have been analytically derived by \cite{zaroubi93, szalay98, szapudi04, matsubara04, papai08}, and tested against both simulations~\citep{raccanelli10} and real data~\citep{samushia11}.
 
\begin{figure}
\begin{center}
\includegraphics[width=0.7\linewidth]{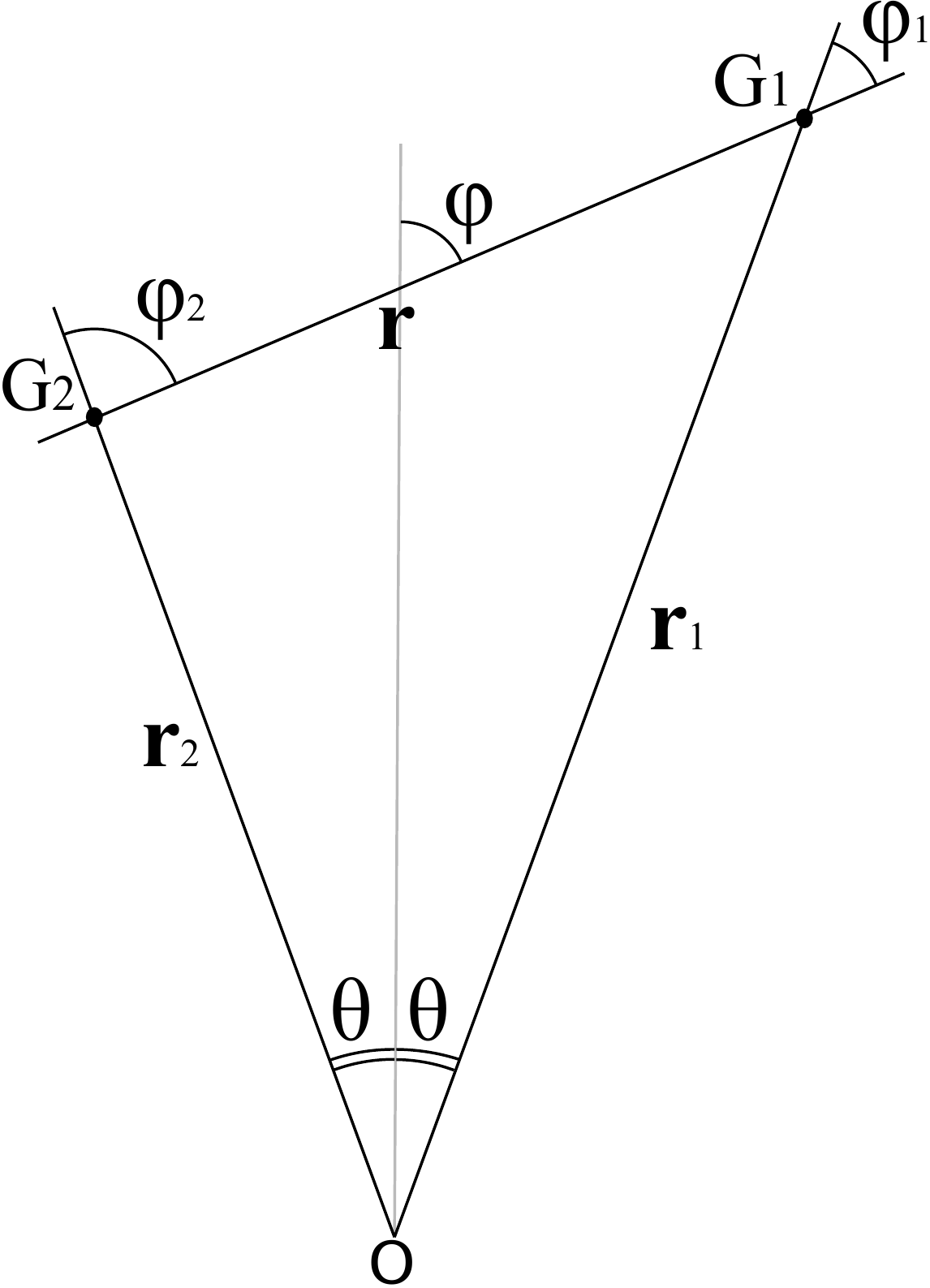}
\caption{{The coordinate system adopted for the triangle formed by the
observer $O$, and galaxies $G_1$ and $G_2$.}}
\label{fig:triangle}
\end{center}
\end{figure}

\citet{papai08} have argued that, for wide angles, the $v/r$ term in Equation~\ref{eq:jacobian} is of the same order as the $\partial_r v$ term.
As a consequence, a careful and precise analysis of the correlation function, requires the full (linear) Jacobian, dropping the distant-observer approximation. In this case, we can express the linear overdensity as:
\begin{equation}  \label{eq:del}
  \delta^s(s) = \int \frac{d^3k}{(2\pi)^3} e^{i k_j \cdot r_j}
    \left[1+f(\mathbf{\hat{r}}_j \cdot \mathbf{\hat{k}}_j)^2 
      - i\alpha(r) f\frac{\mathbf{\hat{r}}_j \cdot \mathbf{\hat{k}}_j}{rk} \right]
    \delta(k);
\end{equation}
the redshift-space correlation function reads:
\begin{align}  
  \xi^s &= 
  \langle \delta^s(\mathbf{s}_1) \delta^{s*}(\mathbf{s}_2) \rangle = \int \frac{d^3k}{(2\pi)^3} P(k) e^{ik(r_1-r_2)} \cdot \nonumber \\ 
  & \left[ 1 + \frac{f}{3} + \frac{2f}{3} L_2(\mathbf{\hat{r}_1} \cdot \mathbf{\hat{k}}) 
    - \frac{i\alpha(r) f}{r_1 k} L_1(\mathbf{\hat{r}}_1 \cdot \mathbf{\hat{k}}) \right] \cdot \nonumber \\
  & \left[ 1 + \frac{f}{3} + \frac{2f}{3} L_2(\mathbf{\hat{r}}_2 \cdot \mathbf{\hat{k}}) 
    + \frac{i\alpha(r) f}{r_2 k} L_1(\mathbf{\hat{r}}_2 \cdot \mathbf{\hat{k}}) \right] .
    \label{eq:xiwa}
\end{align}
The third terms in the brackets describe the wide-angle effects, while the fourth ones are responsible for the so called mode-coupling. The $r_1$ and $r_2$ terms in the denominator depend on the angular separation of the galaxies, and $\alpha$ is proportional to the logarithmic derivatives of the galaxy distribution function (Equation~\ref{eq:alpha}).
We refer to Figure~\ref{fig:triangle} for the geometry of the problem, where we define $2\theta$ to be the angular separation of the two galaxies considered, $\phi_1$ as the angle between the vector to the first galaxy in a pair $\bf{r}_1$ and  $\bf{r}$, $\bf{r}$ to be the vector connecting galaxies in a pair, and $\phi_2$ to be the angle between vector to the second galaxy in a pair $\bf{r}_2$ and $\bf{r}$.

%
Tripolar spherical harmonics are the most natural basis for the expansion of a function that depends on three directions
\citep{varshalovich88}, so, as suggested by \cite{szapudi04} and \cite{papai08}, we expand Equation~\ref{eq:xiwa}
using a subset of them, so that the redshift-space correlation function can then be written as:
\begin{equation}
\xi^s(\hat{r}_1,\hat{r}_2,\hat{r}) = \sum_{\ell_1,\ell_2,\ell} B^{\ell_1\ell_2\ell}(r,\phi_1, \phi_2)S_{\ell_1\ell_2\ell}(\hat{r}_1, \hat{r}_2, \hat{r}), 
\label{eq:xiwaB}
\end{equation}
where $B^{\ell_1\ell_2\ell}(r,\phi_1, \phi_2)$ are a series of coefficients that depend on $f$, $g_i(\phi_i)$ and $\xi^r_{\ell}(r)$ (see~\citealt{raccanelli10} for details on the definition of these functions). 
These coefficients can be divided into two different subsets:
one that depends on the third term inside the brackets in Equation~\ref{eq:xiwa}, given by $B^{\ell_1\ell_2\ell}(r)$, with $\ell_1\ell_2\ell$ combinations of $0$, $2$, $4$, with only the radial dependence accounting for the wide angle effects, 
and one that depends on the fourth terms inside the brackets of Equation~\ref{eq:xiwa}, given by
$B^{\ell_1\ell_2\ell}(r,\phi_1, \phi_2)$, with $\ell_1\ell_2\ell$ combinations of $0$, $1$, $2$, $3$, with also an angular dependence, describing the mode coupling part
(see \cite{szalay98}, \cite{szapudi04} and \cite{papai08} for a detailed derivation). 
The plane-parallel approximation emerges as a limit when $\hat{r}_1=\hat{r}_2$.
This formalism was shown to accurately reproduce wide-angle effects seen in numerical simulations~\citep{raccanelli10}.

\subsection{SDSS DR7}
An extensive analysis of the SDSS DR7 data is provided in \cite{samushia11}; we refer the reader to that paper for the full details on the analysis of SDSS DR7 data and for the relative importance of various corrections considered.
In the following we use the measurements of momenta of the correlation function presented there, to test models of gravity.

\subsubsection{Distribution of $\theta$ and $\mu$}
\label{sec:distributions}
For surveys that cover a significant fraction of the sky, the distribution of galaxies pairs has a complicated dependence on the variables $\{r, \mu, \theta\}$, since not all sets of their combinations are equally likely or even geometrically possible. In particular, the distribution of $\mu$ does not correspond to that of an isotropic pair distribution, and this will strongly bias measurements of angular momenta of the correlation function.
We can see in Section~\ref{sec:appmethod} the effects on the correlation function of a non-zero fixed angular separation, and in Section~\ref{sec:degeneracy} that these errors can bias our estimate of the \gam parameter.

\begin{figure}
\includegraphics[width=\linewidth]{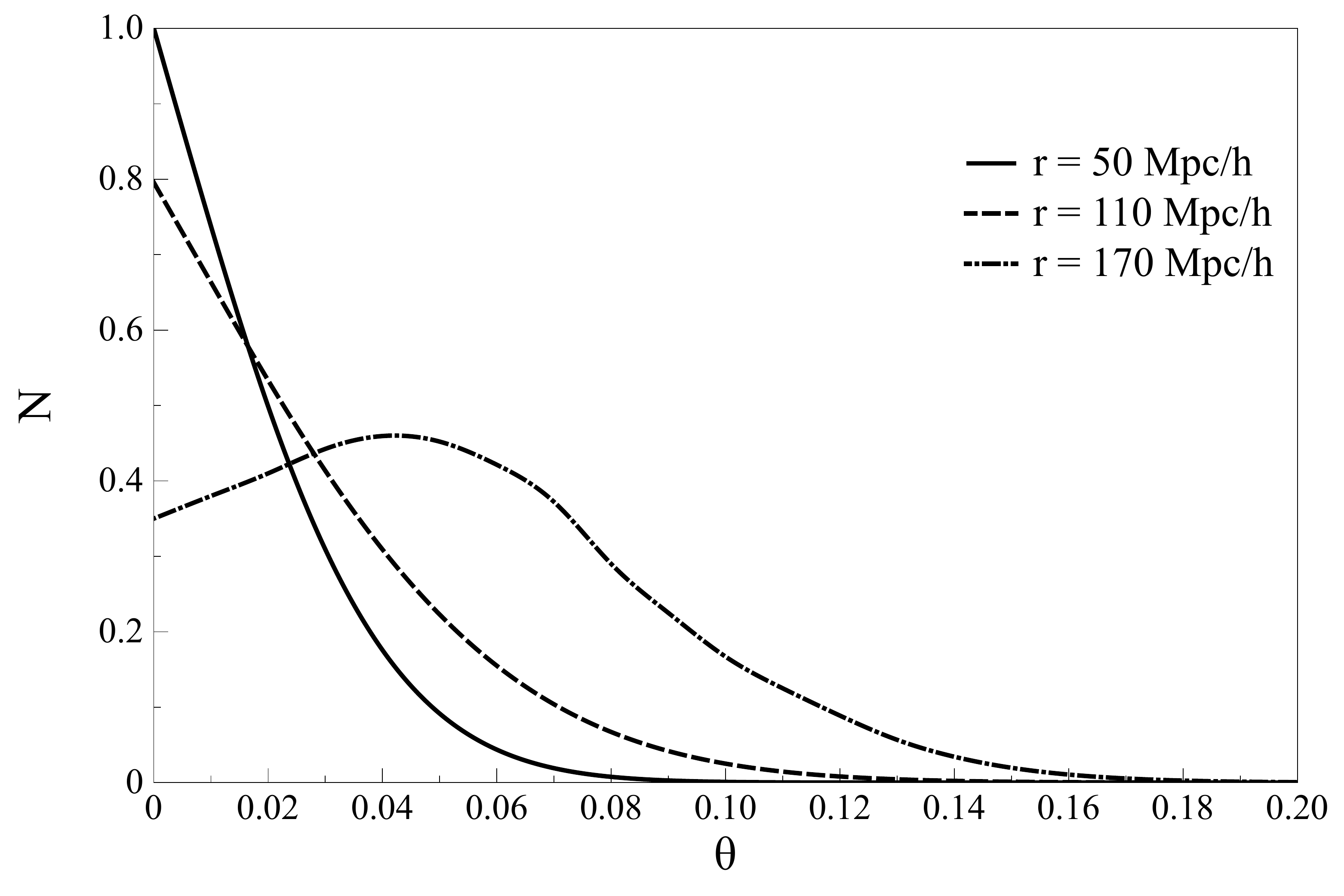}
\caption{Measured distribution of angular separation $\theta$ for LRG pairs in the SDSS DR7 catalog, at different scales.}
\label{fig:theta_dist}
\end{figure}

In Figure~\ref{fig:theta_dist} is shown the distribution of $\theta$ of observed LRGs in the SDSS DR7 catalog.
As expected, in this case the number of pairs with relatively large $\theta$ is very small, and it increases with the linear separation of pairs.

\begin{figure}
\includegraphics[width=\linewidth]{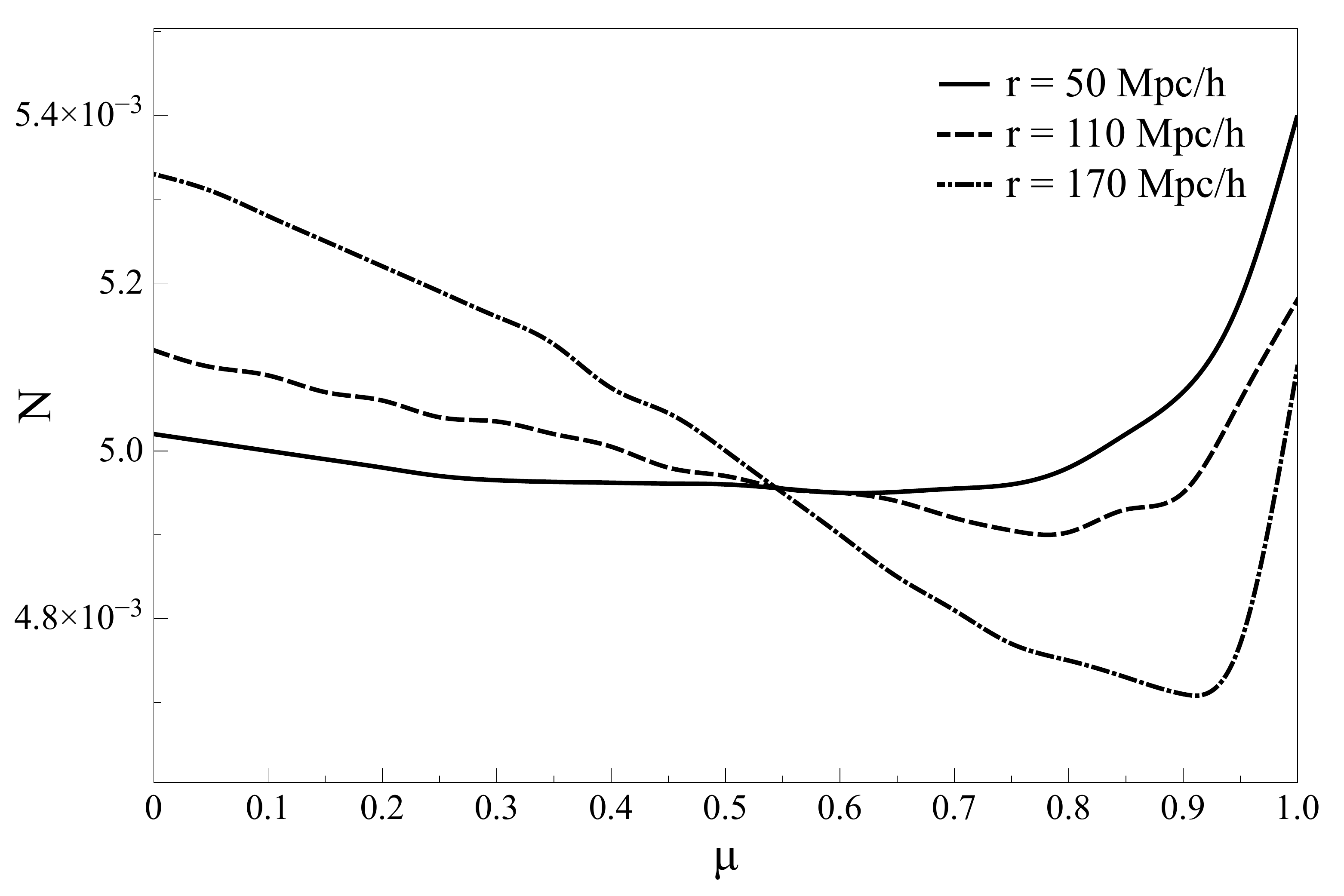}
\caption{Measured distribution of orientation angle $\mu$ for LRG pairs in the SDSS DR7 catalog, at different scales.}
\label{fig:mu_dist}
\end{figure}

Other than the wide-angle and the mode coupling corrections, there is also another difference that has to be taken in account when doing a real-data wide-angle analysis, that derives from the fact that  the distribution of galaxies in $\mu$ will be non-trivial, with some values of $\mu$ not permitted for non-zero $\theta$. As a consequence, we will not be able to measure pure Legendre momenta of the correlation function, but instead we will need to use weighted integrals and biased momenta; corrections due to a non-uniform ``$\mu$-distribution'' can be applied to both plane-parallel and wide-angle analyses.
In Figure~\ref{fig:mu_dist} is shown the distribution of $\mu$ for observed LRGs in the SDSS DR7 catalog.

\subsubsection{Non-linear Effects}
\label{sec:nl}
We model two non-linear effects: one due to the Baryonic Acoustic Oscillations (BAO) peak and the other due to the so-called Fingers of God. Since we are only interested in the signal on large-scales, where the linear theory is an accurate description, we assume that the non-linear effects are small, except for the fact that the matter power spectrum itself goes non-linear.
We describe the non-linear contribution of the power spectrum by means of a two-component model, which splits $P(k)$ into a ``smooth" part that describes the overall shape and a ``wiggled'' part, which describes the BAO:
\begin{equation}  
\label{eq:pwiggled}
  P_{\rm BAO}(k,\mu) = P_{\rm full}(k,\mu)-P_{\rm smooth}(k,\mu) \, .
\end{equation}

The primary non-linear effect on the BAO component of the power spectrum is a damping on small scales, which can be well approximated by a Gaussian smoothing \citep{bharadwaj96, crocce06, crocce08, eisenstein07, matsubara08a, matsubara08b, Matarrese:2007wc}:
\be
\label{eq:nlbao}
  P^{\rm nl}_{\rm BAO}(k,\mu) = P^{\rm lin}_{\rm bao}(k,\mu) \exp\left\{-k^2\left[\frac{(1-\mu^2)\Sigma_\bot}{2}+\frac{\mu^2\Sigma_{||}}{2}\right]\right\},
\ee
where $\Sigma_\bot=\Sigma_0D$ and
$\Sigma_{||}=\Sigma_0(1+f)D$; $\Sigma_0$ is a constant
phenomenologically describing the diffusion of the BAO peak due to
non-linear evolution. From N-body simulations its numerical value is of order
10 Mpc/h and seems to depend linearly on $\sigma_8$ but only weakly on $k$ and other cosmological parameters. \\

Within dark matter haloes the peculiar velocities of galaxies are highly
non-linear. These velocities can induce RSD that are larger than the real-space
distance between galaxies within the halo. This gives rise to the observed
fingers of god (FOG) effect, that is a strong elongation of structures along the line of
sight~\citep{jackson72}. The FOG effect sharply reduces the power
spectrum on small scales compared to the predictions of the linear model, and is
usually modeled by multiplying the linear power-spectrum by a function $F$, that depends on the average velocity
dispersion of galaxies within the relevant haloes, $\sigma_{v}$, $k$ and $\mu$.
The most common one is a Gaussian damping \citep[see e.g.][]{cole95, peacock96}:
\begin{equation}
F(\sigma_{ v},k,\mu) = \exp\left[-(k\sigma_{v}\mu)^2\right] \, ,
    \label{eq:F_gauss}
\end{equation}
that is small on small scales and approaches unity for scales larger than
$1/\sigma_{ v}$.

\subsection{The Estimator of the Correlation Function Momenta}
To estimate the momenta of the correlation function we use Landy-Szalay type estimators \citep{landy93}:
\begin{align}  
\label{eq:ls_ell}
&\hat{\xi_\ell}(r_i) = L_\ell(\mu) \cdot \\ 
& \cdot \displaystyle \frac{\sum_{j,k}
  \left[DD(r_i,\mu_j,\theta_k)-2DR(r_i,\mu_j,\theta_k)
  +RR(r_i,\mu_j,\theta_k)\right]}{\sum_{j,k}\left[RR(r_i,\mu_j,\theta_k)\right]} , \nonumber
\end{align}
where $\mu$$=$${\rm cos}({\varphi})$, while $DD(r_i,\mu_j,\theta_k)$, $DR(r_i,\mu_j,\theta_k)$ and $RR(r_i,\mu_j,\theta_k)$ are the number of galaxy-galaxy, galaxy-random and random-random pairs in bins centered on $r_i$,
$\mu_j$ and $\theta_k$.

RSD measurements are often extracted from the normalised quadrupole $Q$ \citep{hamilton92}, defined as:
\begin{equation}
  \label{eq:Q}
  Q(r) =
  \frac{\xi_2(r)}{\xi_0(r)-\frac{3}{r^3}\displaystyle\int_0^r\xi_0(r')r'^2dr'}.
\end{equation}
The normalised quadrupole $Q$ was introduced because it is independent of the shape of the power spectrum, and so it depends only on the $\beta$ parameters, allowing to directly test gravity; however, this is true only in the Kaiser analysis, and so it is not true in our case. For this reason we test different cosmological models fitting the momenta of the correlation function, $\xi_0, \xi_2$.

We compute error bars as the square root of the diagonal terms in the covariance matrix:
\begin{equation}  
\label{eq:cstat}
{\bf C} = \frac{1}{79}\displaystyle\sum 
\left[\hat{\bf X}(r_i)-{\bf \overline{X}}(r_i)\right]
\left[\hat{\bf X}(r_j)-{\bf \overline{X}}(r_j)\right],
\end{equation}
where $\hat{\bf X}(r)$ is a vector of the measurements of $\xi_0, \xi_2$ at scale $r$ and ${\bf \overline{X}}$ is the mean value from all 80 mock catalogues.

We compare measurements with predictions from different models, where we compute the redshift-space correlation function including effects from wide-angle and $\mu$-distribution as well as survey geometry and non-linearities.
Estimates of Legendre momenta given by Equation~\ref{eq:ls_ell} correspond to:
\begin{equation}
  \tilde{\xi}_{\rm \ell}(r) 
  = \displaystyle\int W_{\rm r}(r, \theta, \mu) \, \xi^s(r,\theta,\mu) \, L_{\rm \ell}(\mu) \, d\theta \, d\mu,
\end{equation}
\noindent
where $W_{\rm r}(r, \theta, \mu)$ is a weight function that appears because of the geometrical constraints and the not uniform distribution of $\mu$, as explained in Section~\ref{sec:warsd} and gives the relative number of pairs in
a survey that form angles $\mu$ and $\theta$ for a given scale $r$; when the $\theta$ distribution tends towards a delta function centred at $\theta=0$, the wide-angle effects become negligible, while when the distribution in $\mu$ tends towards a uniform one, this effect becomes negligible. 
The correlation function of Equation~\ref{eq:xiwaB} can be written as:
\begin{equation}
\label{eq:xiwa_cab}
  \xi^s(r,\theta,\varphi) = \sum_{ab} c_{ab}(f,r) 
    L_a[\cos(\theta)] L_b[\cos(\varphi)] ,
\end{equation}
where $\{r,\theta,\varphi\}$ are as in Fig.~\ref{fig:triangle}, 
and $c_{ab}(f,r)$ are coefficients that depend on the separation between galaxies and the $f$ of Equation~\ref{eq:flogd} (see~\citealt{szapudi04, papai08, raccanelli10} for further details).

\subsection{Las Damas Mocks}
\begin{figure}
\includegraphics[width=\linewidth]{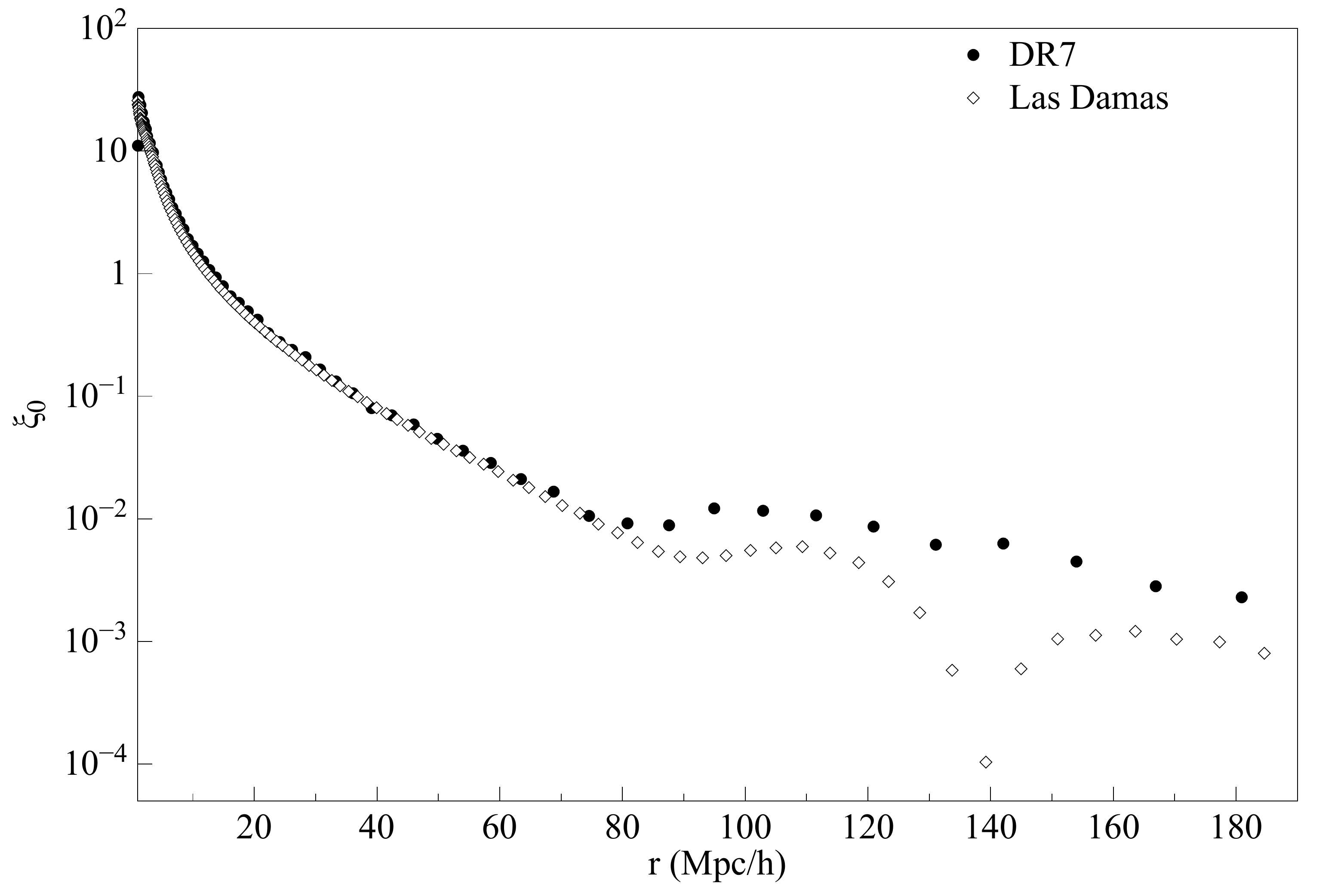}
\caption{$\xi_0$ measured from 80 Las Damas mock catalogs and from SDSS DR7 data.}
\label{fig:xi0meascomp}
\end{figure}

\begin{figure}
\includegraphics[width=\linewidth]{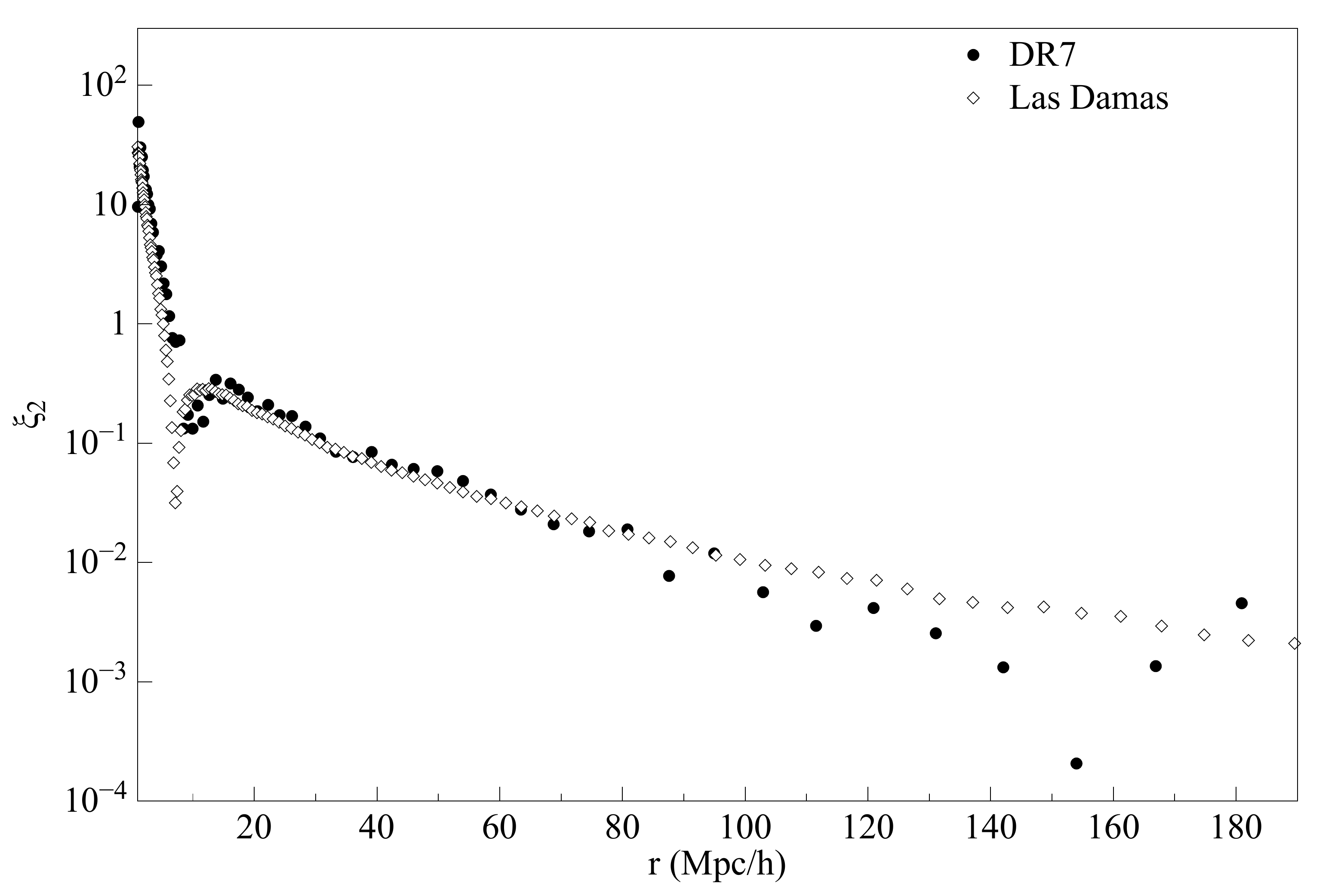}
\caption{$\xi_2$ measured from 80 Las Damas mock catalogs and from SDSS DR7 data.}
\label{fig:xi2meascomp}
\end{figure}

In this paragraph we show our measurements from the LasDamas simulations and the \lcgr model prediction.
As one can see in Figures \ref{fig:xi0meascomp}, \ref{fig:xi2meascomp}, the LasDamas set reproduces quite well the measured momenta of the correlation function, the main deviation being on large scales of the monopole of the correlation function, that represents also the main source of deviation from \lcgr at the scales we considered; however the other models don't help much in picking up that deviation.
This is a well known discrepancy between theoretical predictions and observations, and it has been detected in spectroscopic (\citealt{kazin10}, \citealt{samushia11}) and photometric (\citealt{thomas10}) data sets.
This excess power at large scale can be induced by primordial non-Gaussianity (see e.g. \citealt{matarrese2008, dalal08, slosar08, desjacques10, xia10}) or exotic physics \citep{thomas10}; however \citealt{samushia11} found that the redshift dependence of the excess power is different from what would be caused by non-Gaussianity, and \citet{ross11} suggest that this is likely to be due to masking effects from stellar sources.
After correcting for systematics, \citet{ross12} found consistency at better than $2\sigma$ between the BOSS CMASS DR9 large-scale clustering data and the WMAP LCDM cosmological model. Further tests using mocks suggested that there was no evidence that additional potential systematic trends contributed individually at a level above that expected through noise. But, there is always the potential for more systematic problems to be present, or for the combination (which always tends to add power) not to have a more-significant contribution - i.e. form part of the $2\sigma$ discrepancy.

\begin{figure}
\includegraphics[width=\linewidth]{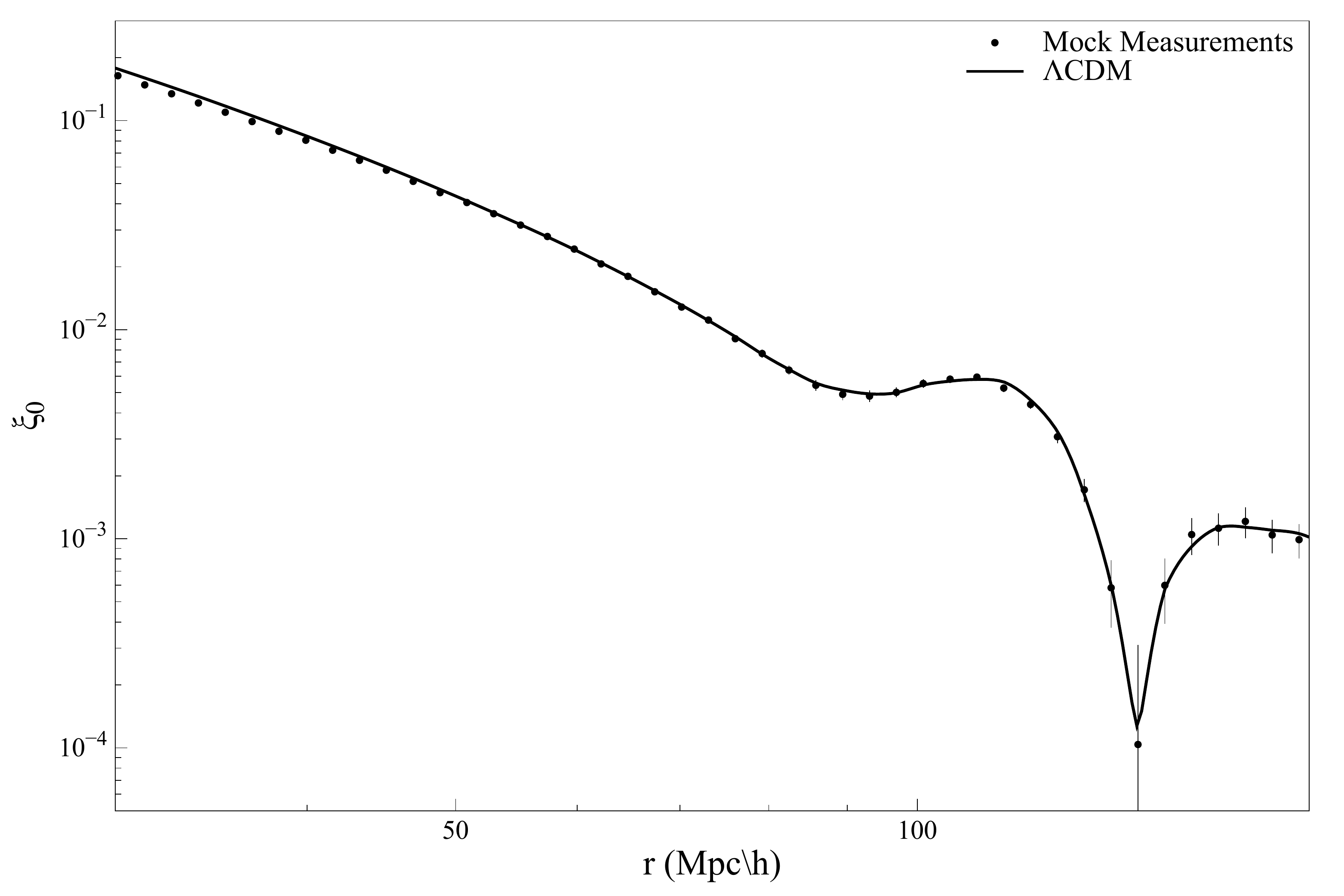}
\caption{$\xi_0$ measured from 80 Las Damas mock catalogs and theoretical predictions for \lcgr.}
\label{fig:xi0mocks}
\end{figure}

\begin{figure}
\includegraphics[width=\linewidth]{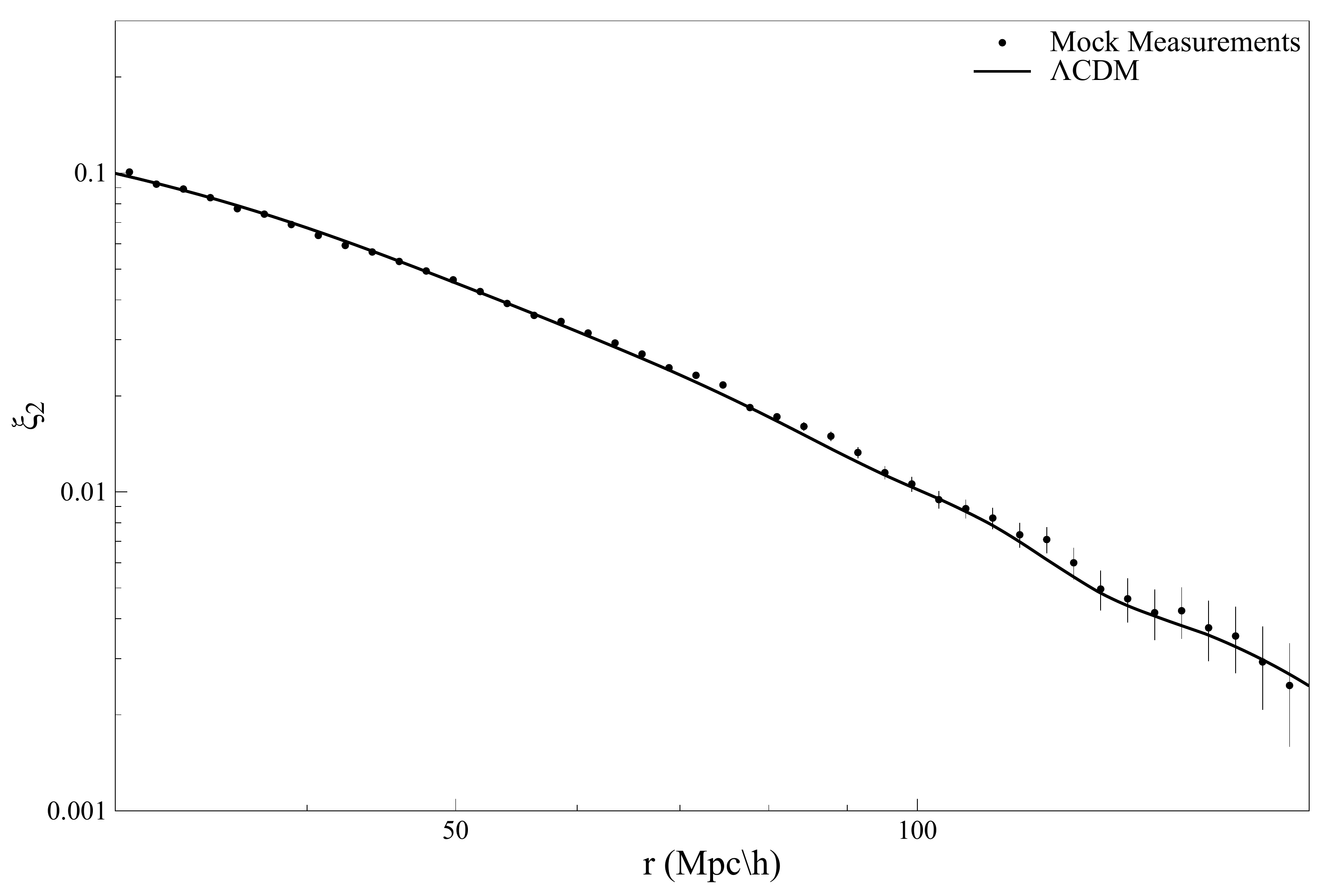}
\caption{$\xi_2$ measured from 80 Las Damas mock catalogs and theoretical predictions for \lcgr.}
\label{fig:xi2mocks}
\end{figure}

Figures \ref{fig:xi0mocks}, \ref{fig:xi2mocks} show that our methodology, that includes corrections due to non-linearities, wide-angle and $\mu$-distributions, can fit the \lcgr simulations very well, up to scales of 180 Mpc/h;
For the data analyses, however, we decided to use as a maximum scale r = 120 Mpc/h.

\end{document}